\documentclass[numbers,webpdf]{ima-authoring-template}%
\usepackage{amsmath,amssymb,amsfonts}
\graphicspath{{Fig/}}

\usepackage{calrsfs}
\DeclareMathAlphabet{\pazocal}{OMS}{zplm}{m}{n}
\newcommand{\red}[1]{\textcolor{black}{#1}}

\newcommand{\beq}{\begin{equation}}
\newcommand{\eeq}{\end{equation}}
\newcommand{\pfr}[2]{\ensuremath{\frac{\partial #1}{\partial #2}}}
\newcommand{\pfi}[2]{\ensuremath{{\partial #1}/{\partial #2}}}
\newcommand{\ep}{\varepsilon}
\newcommand{\tr}{\mathrm{tr}}
\newcommand{\mb}[1]{\mathbf{#1}}
\newcommand{\gb}[1]{\boldsymbol{#1}}
\newcommand{\mc}[1]{\pazocal{#1}}

\begin{document}

\DOI{DOI HERE}
\copyrightyear{2025}
\vol{00}
\pubyear{2021}
\access{Advance Access Publication Date: Day Month Year}
\appnotes{Paper}
\copyrightstatement{Published by Oxford University Press on behalf of the Institute of Mathematics and its Applications. All rights reserved.}
\firstpage{1}


\title[Colloidal nanoparticles in liquid crystals]{Colloidal nanoparticles in liquid crystals: Bulk properties, biaxiality and untwisting in cholesterics}

\author{Prabakaran Rajamanickam\ORCID{0000-0003-1240-0362}
\address{\orgdiv{Department of Mathematics and Statistics}, \orgname{University of Strathclyde}, \orgaddress{\street{Glasgow}, \postcode{G1 1XQ},  \country{UK}}}}

\author{Fatimah Almutari
\address{\orgdiv{Department of Mathematical Sciences, College of Science}, \orgname{Princess Nourah bint Abdulrahman University, P.O. Box 84428}, \orgaddress{\street{Riyadh}, \postcode{11671},  \country{Saudi Arabia}}}
\address{\orgdiv{Department of Mathematics and Statistics}, \orgname{University of Strathclyde}, \orgaddress{\street{Glasgow}, \postcode{G1 1XQ},  \country{UK}}}}

\author{Apala Majumdar\ORCID{https://orcid.org/0000-0003-4802-6720}
\address{\orgdiv{Department of Mathematics}, \orgname{University of Manchester}, \orgaddress{\street{Manchester}, \postcode{M13 9PL},  \country{UK}}}}

\authormark{Rajamanickam, Almutari and Majumdar}


\received{Date}{6}{2025}
\revised{Date}{0}{2025}
\accepted{Date}{0}{2025}


\abstract{We study the effects of colloidal nanoparticles (NPs) in liquid crystal samples in the dilute limit, in a Landau--de Gennes theoretical framework. The effects of the suspended NPs are captured by a homogenized energy, as outlined in~\cite{canevari2020design}. For spatially homogeneous samples, we explicitly compute the critical points and minimizers of the modified Landau--de Gennes energy and show that the presence of NP eliminates the first-order isotropic-nematic phase transition, stabilises elusive biaxial phases over some temperature ranges and that the symmetry of the NP boundary conditions or surface treatments dictates the bulk equilibrium phase at high temperatures. We also numerically demonstrate structural transitions from twisted helical director profiles to untwisted director profiles in cholesteric-filled channel geometries, driven by the collective effects of the NPs and increasing temperature. These transitions are reversible upon lowering the temperature in sufficiently large domains, where thermal hysteresis can also be observed. This behaviour opens interesting avenues for tuning the optical properties of confined, nano-doped cholesteric systems.}


\maketitle

\section{Introduction}
\label{sec:intro}

Liquid crystals (LCs) are paradigm examples of partially ordered materials that combine fluidity with some of the ordering characteristics of solids~\cite{dg}. LC phases exhibit some form of long-range orientational order, i.e., the LC molecules tend to align along certain locally preferred directions, referred to as \emph{directors} and consequently, LCs have direction-dependent physical and rheological properties~\cite{dg}. This intrinsic directionality or anisotropy renders LCs  to be exceptionally responsive and functional soft materials, with widespread applications across the physical sciences and engineering \cite{lagerwallnew, bisoyi}. 

There are three canonical LC phases: the simplest nematic (NLC) phase with no positional ordering but with long-range orientational ordering characterised by \emph{nematic directors}; the smectic phase with orientational and partial positional ordering, i.e., the smectic molecules arrange themselves in layers that are stacked on top of each other and the cholesteric (CLC) phase, which is similar to the NLC phase except that the cholesteric director naturally twists in space, rendering stable helical director profiles with exceptional optical properties~\cite{dg}. NLCs are the working material of choice for a variety of electro-optic devices, e.g., the multi-billion dollar liquid crystal display industry, largely because of their strong responses to light and electric fields. CLCs exhibit distinctive helical director profiles and their natural twistedness, or chirality, has found many uses in photonics, sensors, anti-counterfeiting and bio-inspired systems ~\cite{lagerwallnature, mitov2017cholesteric}. There is substantial contemporary interest in doping NLC and CLC systems with colloidal nanoparticles (NPs) to tune their optical and mechanical properties by means of NP shapes, sizes, distributions and surface treatments on NP-LC interfaces \cite{kumar2018quantum, mirzaei2011nanocomposites, mitov2017cholesteric, kinkead2010effects, zhang2009cds}. For example, in \cite{patranabish2021quantum}, the authors experimentally and theoretically investigate a dilute suspension of quantum dots in a bent-core nematic system, presenting detailed experimental results on optical textures, phase transition temperatures and dielectric anisotropy data of these doped systems. These quantum dots, typically synthesised using Cd and Zn-based compounds, are capped with suitable ligands to prevent aggregation, making them good candidates for suspension in the LC matrix~\cite{patranabish2021quantum}. 
The authors demonstrate that the quantum dots and, NPs in general, can be used to tune the dielectric anisotropy, phase transition temperatures and  optical textures of confined bent-core nematic systems~\cite{patranabish2021quantum}; related works can also be found in \cite{basu2009evidence,mirzaei2011nanocomposites,mirzaei2012quantum,singh2016emissivity,kumar2018quantum} with regards to the effects of NPs on operating voltages, elastic constants and response times in LC systems. In other words, doping offers exciting prospects for tailor-made or customised LC systems on demand, as well as tunable metamaterials in general.

In this paper, \red{we study a prototype system comprising a dilute suspension of colloidal spherical NPs in a LC sample (nematics or cholesterics), although many of our conclusions will naturally extend to generic LC systems.} In a dilute suspension, the NP size is much smaller than the average distance between NPs, which in turn is much smaller than the macroscopic sample length scales. The total volume fraction of the NPs is much less than unity and in this dilute limit, the collective effects of the NPs can be effectively homogenized into a \emph{bulk effect}. We work in the celebrated Landau--de Gennes (LdG) theoretical framework for NLCs and CLCs, wherein the macroscopic NLC or CLC state is described by the macroscopic $\mb Q$-tensor order parameter \cite{dg,mottram2014introduction,majumdar2010equilibrium}. The $\mb Q$-tensor order parameter is relatable to experimentally measurable quantities such as dielectric anisotropy and magnetic susceptibility, which characterize the LC's response to external electric and magnetic fields \cite{dg}. Mathematically speaking, the LdG $\mb Q$-tensor is a symmetric, traceless $3 \times 3$ matrix, whose eigenvectors model the LC directors, i.e., the special directions of averaged molecular alignment, and whose corresponding eigenvalues are \emph{scalar order parameters} that quantify the degree of orientational ordering. A LC phase is uniaxial if it has a single distinguished material direction and  is biaxial if it has two special directions; more details are given in the next section. The LdG theory has an associated free energy, typically of the form
\[
\mc F[\mb Q]:= \int \{f_B(\mb Q) + f_{el}(\mb Q, \nabla \mb Q)\}~dV
\] 
  in the absence of external fields or surface effects. Physically observable configurations are modelled by the LdG energy minimizers subject to the problem constraints. In the case of spatially homogeneous samples, the LdG energy reduces to the bulk potential, $f_B$, which is typically a polynomial in the scalar invariants of $\mb Q$ \cite{majumdar2010equilibrium}. The bulk equilibrium phases are then modelled in terms of the minimizers of $f_B$, which can be computed explicitly as a function of temperature \cite{mottram2014introduction}. The bulk potential, $f_B$, is identical for NLC and CLC phases. The elastic energy density, $f_{el}$, depends on both $\mb Q$ and its gradients, and penalises spatial inhomogeneities. In fact, $f_{el}$, has different forms for the NLC and CLC phases because the natural ground state for a CLC is a helical director profile whereas that for a NLC is a uniform, constant director profile. The critical points and the minimizers of $\mc F$ are classical solutions to the associated Euler-Lagrange equations: a system of nonlinear, coupled partial differential equations, subject to prescribed constraints.

The question of interest is: how is the LdG energy modified by the NPs in the dilute limit described above? This question has been answered in \cite{canevari2020design, canevari2020polydispersity}. In the dilute limit, the NPs do not interact with each other and the NP-LC interaction is captured by a surface energy, $f_s$, that depends on the NP geometry and the surface treatment on the NP-LC interface, $\Sigma$. The collective effects of the NPs is, informally speaking, the sum of the surface energies over all NP-LC interfaces which yields a homogenized bulk energy, $f_{hom}(\mb Q)$. More precisely, the total LdG energy, including the effects of the suspended NPs is given by
\[
\mc F_{\epsilon}[\mb Q]: = \mc F[\mb Q] + \epsilon^{\gamma} \sum_{i}\int_{\Sigma} f_s\left(\mb Q, \mb Q_\nu, \gb \nu \right)~dS
\]
where $\epsilon$ is related to the NP size, $\gamma$ is an appropriate positive scaling factor determined by the dilute limit, $\nu$ is the normal to the NP-LC interface, $\mb Q_\nu$ is a preferred boundary condition on $\Sigma$ and we sum the surface energies over all suspended NPs. In \cite{canevari2020design}, the authors show that minimizers of $\mc F_\epsilon$ converge to minimizers (strongly in $H^1$) of
\[
\mc F_0[\mb Q]: = \int\{ f_B(\mb Q) + f_{el}(\mb Q, \nabla \mb Q) + f_{hom}(\mb Q)\}~dV
\]
as $\epsilon \to 0$ (which captures the dilute limit) and $f_{hom}$ can be explicitly computed in terms of the NP shape, distribution and NP boundary treatments.

We apply this homogenization framework to a \red{dilute suspension of colloidal NPs in a NLC/CLC sample} and analyse the properties of the NP-doped systems in terms of the minimizers of $\mc F_0$, as introduced above. In Section~\ref{sec:homo}, we compute the bulk equilibrium phases of these NP-doped systems as functions of the NP properties and  temperature. The bulk phases are modelled by the minimizers of the modified bulk potential, which is the sum of $f_B$ and $f_{hom}$. \red{In fact, the bulk equilibrium phases do not distinguish between NLC and CLC samples and hence the conclusions in Section~\ref{sec:homo} apply to both NLC and CLC samples.} The NP properties are captured by two parameters: an anchoring coefficient, $W$, on the NP-LC interfaces and a phenomenological matrix, $\mb X$, that can be computed from $f_s$ directly. We find that NPs stabilise elusive biaxial phases in certain temperature ranges, which are typically inaccessible in pure NLC or CLC samples. Equally importantly, the collective symmetry of the NPs (modelled by $\mb X$) has a direct bearing on the bulk equilibrium phase at high temperatures and breaks the inherent rotational symmetries of the NLC and CLC phases. These concepts are beautifully illustrated by means of phase diagrams in Section~\ref{sec:homogeneous}, which numerically capture how the bulk equilibrium phases depend on $\mb X$, $W$ and temperature. In Section~\ref{sec:inhomogeneous}, we apply this framework to a spatially inhomogeneous CLC-filled channel geometry, doped with spherical NPs in the dilute limit. In the absence of NPs, CLCs naturally relax into twisted helical profiles for relatively large channel widths. The NPs induce untwisting transitions in these geometries as temperature increases, i.e., the NPs naturally break the helical symmetry of the CLC phase. In other words, the NPs collectively impose a special direction (determined by $\mb X$) on the CLC phase, for sufficiently high temperatures and large channel geometries. These temperature-induced untwisting transitions are specific to CLCs. However, using the same arguments, we do expect NPs to induce symmetry-breaking transitions in other confined LC systems too. We conclude with some perspectives in Section~\ref{sec:conclusions}.

\section{Homogenized bulk energy}
\label{sec:homo}

We study a dilute system of suspended colloidal NPs, e.g., quantum dots, in a \red{nematic/cholesteric liquid crystal-filled channel geometry} defined by
\begin{equation}
    \label{eq:omega}
\Omega := \left\{\mb r =(x,y,z); 0\leq x\leq L, 0\leq y \leq D, 0\leq z \leq H \right\},
\end{equation}
with $H \ll L\sim D$. We work in the celebrated phenomenological Landau--de Gennes (LdG) framework~\cite{de1971short,dg}. The LdG framework describes the macroscopic LC state in terms of the LdG $\mb Q$-tensor: a symmetric, traceless $3 \times 3$ matrix that belongs to the space
\begin{equation}
    \mb S :=  \{\mb Q \in \mathbb R^{3\times 3}; \mb Q= \mb Q^T, \tr\,\mb Q =0\}. \label{eq:1}
\end{equation}
We often refer to the eigenvector of $\mb Q$ with the largest positive eigenvalue as the \emph{director}, i.e., the molecules tend to align along the director or exhibit long-range orientational ordering along the director, in an averaged sense.

\red{The LC phase is said to be isotropic when $\mb Q = 0$ or equivalently, when the eigenvalues of $\mb Q$ are zero and there are no special directions/directors.} 
In the uniaxial phase, $\mb Q$ has two degenerate non-zero eigenvalues and can be written in the form
\begin{equation} \label{eq:uniaxial}
\mb Q_u = s(\mb n \otimes \mb n - \tfrac{1}{3} \mb I)
\end{equation}
where $\mb n$ is the distinguished eigenvector with the non-degenerate eigenvalue and $s$ is a scalar order parameter that measures the degree of orientational order about $\mb n$. The biaxial phase is the most generic phase and occurs when $\mb Q$ has three distinct eigenvalues and two distinguished material directions.

As with most variational theories in materials science, the \red{nematic/cholesteric phase has an associated LdG free energy} and the equilibrium or physically observable configurations are modelled by LdG energy minimizers, subject to  the imposed boundary conditions \cite{majumdar2010equilibrium}. As mentioned in the Introduction, in the absence of any external fields or surface energies, the LdG energy has two essential contributions:
\begin{equation}\label{eq;ldgenergy}
\mc F[\mb Q]:= \int_\Omega \left\{f_B(\mb Q) + f_{el} (\mb Q, \nabla \mb Q)\right\}~dV
\end{equation}
where $f_B$ is the bulk potential, which depends only on the eigenvalues of the $\mb Q$-tensor and $f_{el}$ is the elastic energy density, which depends on both $\mb Q$ and its gradient $\nabla \mb Q$. The bulk potential determines the LC phase (isotropic, uniaxial, or biaxial) in spatially homogeneous samples, as a function of temperature for thermotropic systems. The elastic energy density penalises spatial inhomogeneities stemming from boundary effects, geometric frustration and topological constraints.

In this paper, we work with the simplest form of $f_B$ that allows for a first-order isotropic-nematic phase transition as a function of the temperature. This form of $f_B$ is simply a quartic polynomial in $\mb Q$, as given by \cite{majumdar2010equilibrium}:
\begin{equation}
f_B(\mb Q) = \frac{A}{2}\textrm{tr}\mb Q^2 - \frac{B}{3}\textrm{tr}\mb Q^3 + \frac{C}{4}\left(\textrm{tr} \mb Q^2 \right)^2
\label{eq:2}
\end{equation} 
where $A=\alpha (T - T^*)$ is a rescaled temperature, and $B$ and $C$ are positive material-dependent constants. The units of $A, B, C$ are $N m^{-2}$. Here, $\alpha$ is positive and $T^*$ is a critical temperature such that the isotropic phase becomes unstable for $A<0$. As will be shown in the next section, the critical points of $f_B$ can be explicitly computed as a function of $T$. The minimizer of $f_B$ is the isotropic phase, $\mb Q=0$, at high temperatures, whereas it is an ordered uniaxial phase for low temperatures. \red{Note that the bulk energy density is identical for nematic and cholesteric liquid crystal phases since cholesterics can be viewed, in non-technical terms, as twisted nematics.}

In this section, we build on the methods in \cite{canevari2020design, canevari2020polydispersity} to demonstrate the collective effects of suspended colloidal nanoparticles (NPs) on the bulk potential, in the dilute limit. In other words, we compute the homogenized bulk energy in the dilute limit. 
 There are three effective length scales in the problem: the colloid size $a$, the average interparticle spacing $l$ and the geometric length scales of the domain $\Omega$. In the dilute limit, we assume that the length scales are widely separated, i.e., $a \ll l \ll H \ll L \sim D$; the particle number density $n$ is given by $l=n^{-1/3}$~\cite{batchelor1972sedimentation}, and the volume fraction of the NPs is much smaller than unity $(na^3=(a/l)^3\ll 1)$ so that interparticle interactions can be neglected, provided they are not long-ranged.

These assumptions on the length scales ensure that we have a large number of NPs within each physically infinitesimal volume for effective averaging. The probability distribution of the colloidal particles is effectively described by the one-particle distribution function $p(\mb x, \gb \theta)$, where $\mb x$ is the position vector of a single NP and $\gb\theta$ denotes its three Euler angles with respect to a fixed coordinate system. The distribution function is normalized such that $\int_V\int_\theta p(\mb x,\gb\theta) d\gb\theta dV = N$ (the total number of NPs) with the local number density given by $n(\mb x)=\int_{\theta} p(\mb x,\gb\theta) d\gb\theta$.

In the homogenization problem, the LC molecules are much smaller than the NPs and the NP-LC interaction can be described by a surface energy. The homogenised energy is the collective contribution of the NP-LC surface energies, summed over all the NPs. More precisely, we model the NP-LC interaction energy per unit area by the familiar weak anchoring energy density~\cite{mottram2014introduction},
\begin{equation} \label{eq:3}
    f_s(\mb Q,\mb Q_\nu) = \frac{w}{2}|\mb Q-\mb Q_\nu|^2, 
\end{equation}
where $\mb Q_\nu$ is a prescribed preferred boundary condition on the NP-LC interface, $\Sigma$, and $w$ is an anchoring strength parameter with units $Nm^{-1}$. In the strong anchoring $w \to \infty$ limit, we recover the familiar Dirichlet condition $\mb Q = \mb Q_\nu$ on $\Sigma$.
As demonstrated in prior work using multiple-scale analysis~\cite{bennett2017multiscale,bennett2018multiscale} and mesocharacteristics method~\cite{canevari2020design,canevari2020polydispersity}, the collective influence of the NPs manifests as a bulk energy density at large length scales and is computed to be \cite{canevari2020design, canevari2020polydispersity}:
\begin{align}
    f_{hom}(\mb Q,\mb x) = \frac{w}{2} \int_\theta \int_\Sigma p(\mb x,\gb\theta)|\mb Q(\mb x)- \mc R^T\mb Q_\nu(\gb \xi) \mc R|^2\,d\Sigma d\gb\theta. \label{homo1}
\end{align}
where $\mb x$ is the large-scale coordinate, $\gb\xi$ is the small-scale coordinate that varies on the length scale $a$ and $\mc R=\mc R(\gb\theta)$ is the rotation matrix pertaining to the Euler angles $\gb\theta$. The interface $\Sigma$ is parametrized using $\gb \xi$, with $d\Sigma \sim a^2$.

In this work, we assume that the NPs are uniformly distributed so that $p$ is independent of $\mb x$. Furthermore, we assume that the NPs are spherical in shape and are experimentally subjected to the same boundary conditions, i.e., $\mb Q_\nu$ is the same for all spherical NPs and consequently, $p$ is also independent of $\theta$. Following~\cite{bennett2017multiscale,bennett2018multiscale}, $p(\mb x, \theta) = p$ is a constant under these assumptions and the rotation matrix is simply the identity matrix in~\eqref{homo1}. Expanding the integrand in~\eqref{homo1} then yields
\begin{equation}
    f_{hom}(\mb Q) = \frac{\Sigma_s}{2}na^2 w\,\tr\,\mb Q^2 -  na^2w\, \tr\,(\mb Q\mb X)  + \text{constant}, \label{homo2}
\end{equation}
where $\Sigma_s$ is the area of the NP-LC interface measured in units of $a^2$ (e.g. for a sphere of radius $a$, $\Sigma_s=4\pi$) and
\begin{equation}
    \mb X =  \frac{1}{na^2}\int_\Sigma   \mb Q_\nu\, d\Sigma . \label{Xtensor}
\end{equation}
In general, we can treat the $\mb X$-tensor as a phenomenological $3\times 3$, symmetric, traceless tensor, that is purely determined by the NP-LC surface interactions and represents the collective influence of the spherical NPs, even without knowledge of the underlying microscopic details. For example, if $\mb Q_\nu$ corresponds to homeotropic or radial boundary conditions on a spherical particle, then $\mb X = 0$ \cite{canevari2020design}. For arbitrary $\mb Q_\nu$ and spherical NPs, $\mb X$ can be non-trivial and we consider multiple possibilities for $\mb X$ in the next sections. 

Combining the homogenized NP-LC interaction energy density from~\eqref{homo2} with the standard bulk potential $f_B$ from \eqref{eq:2}, the total modified bulk energy density, $f_m$, is defined to be
\begin{equation}
    f_m(\mb Q) = C\left[\frac{\mc A+\Sigma_s W}{2} \tr\,\mb Q^2 - \frac{ \mc B}{3}\tr\,\mb Q^3 + \frac{1}{4}(\tr\,\mb Q^2)^2  - W \tr(\mb Q\mb X) \right] + \text{constant}, \label{modified}
\end{equation}
where $\mc A = A/C$, $\mc B = B/C$ and $W= na^2w/C$. 
\red{As noted in~\cite{canevari2020design,canevari2020polydispersity},  $f_m$ differs from $f_b$ in two ways - a modified coefficient of $\tr \mb Q^2$ and the addition of a linear forcing term $\tr(\mb Q\mb X)$. The modified coefficient of $\tr \mb Q^2$ implies that colloidal NPs change the effective temperature of the LC system in the dilute limit. This might be attributed to the effects of the NP-LC interactions which induce some heating or cooling effects depending on the surface treatments on the NP-LC interfaces. One can sometimes view $\mb X$ as a collective external field and write $\mb X = \mb E\otimes\mb E$. Here, $\mb E \in \mathbb{R}^3$ is an external field and we can redefine $\mb X=\mb E\otimes\mb E-\tfrac{1}{3}\mb I |\mb E|^2$ to ensure that the isotropic part does not enter into the free energy~\eqref{homo2}.} We use the values $B = 3.7 \times 10^{5}$ $N m^{-2}$ and $C = 2.4 \times 10^{5}$ $Nm^{-2}$ from ~\cite{fukuda2010cholesteric} and assuming $w=5\times 10^{-3} Nm^{-1}$, $a=50$ nm and $n=10^{22}$$m^{-3}$, we obtain $\mc B = 1.5$ and $W=0.5$, which will be used in the subsequent numerical computations. 

In the next section, we outline the procedure for computing the critical points of \eqref{modified} with isotropic, uniaxial and biaxial $\mb X$ respectively. We defer the discussion of spatially inhomogeneous systems to  Section~\ref{sec:inhomogeneous}.

\section{Spatially homogeneous systems}
\label{sec:homogeneous}

Consider the bulk potential $f_B$ in \eqref{eq:2}; one can explicitly show that all critical points of $f_B$ are either isotropic or uniaxial $\mb Q$-tensors of the form \eqref{eq:uniaxial} \cite{majumdar2010equilibrium}. In fact, one can explicitly show that there are three critical temperatures associated with the quartic polynomial, $f_B$: (i) $\mc A=0$, below which the isotropic phase is unstable and the minimizers of $f_B$ are a continuum of uniaxial phases denoted by
\begin{equation}
    \label{eq:fbmin}
    \mb Q_{\min} = s_+ ( \mb n \otimes \mb n - \tfrac{1}{3} \mb I)
\end{equation}
where
\begin{equation}
\label{eq:s+}
s_+ = \frac{B + \sqrt{B^2 - 24 A C}}{4 C} = \frac{\mc B + \sqrt{\mc B^2 - 24 \mc A}}{4}
\end{equation} and $\mb n \in S^2$ is arbitrary. (ii) The isotropic-nematic phase transition temperature, $\mc A = \frac{\mc B^2}{27}$ for which $f_B\left(\mb Q_{\min}\right) = f_B(\mb Q = 0)$,  and (iii) $\mc A = \frac{\mc B^2}{24} $, above which $\mb Q = 0$ is the unique minimizer of $f_B$.

As mentioned before, the minimizers and critical points of $f_B$ determine the competing equilibrium/physically observable configurations in spatially homogeneous systems, as a function of temperature. The crucial question is: how do the suspended NPs modify these bulk equilibrium phases, that are modelled by the minimizers of the modified bulk potential~\eqref{modified}? The critical points (including minimizers) of $f_m$ are solutions of the following system of five algebraic equations (subject to the constraint $\tr\,\mb Q=0$):
\begin{equation}
   (\mc A + \Sigma_s W) \mb Q - \mc B \left(\mb Q^2 - \tfrac{1}{3}\mb I\,\tr\,\mb Q^2 \right) +  \mb Q\, \tr\,\mb Q^2  = W \mb X, \label{bulkEL}
\end{equation} where $\mb X$ is a known matrix from \eqref{Xtensor}.
An elementary calculation shows that any critical point or solution $\mb Q$ of \eqref{bulkEL} has the same set of eigenvectors as $\mb X$. Since $\mb Q$ is symmetric, there exists an $\mb R\in SO(3)$ such that $\mb D = \mb R^T \mb Q \mb R$ is a diagonal matrix. Transforming~\eqref{bulkEL} using $\mb R$ yields
\begin{equation}
   (\mc A + \Sigma_s W) \mb D - \mc B \left(\mb D^2 - \tfrac{1}{3} \mb I\,\tr\,\mb Q^2 \right) +  \mb D\, \tr\,\mb Q^2 = W \mb R^T \mb X \mb R.
\end{equation}
The left-hand side of the above equation is a diagonal matrix, and hence, $\mb R^T\mb X \mb R$ must also be a diagonal matrix. Consequently, $\mb Q$ and $\mb X$ possess the same set of eigenvectors.

Since the frame of reference is arbitrary for spatially homogeneous systems, we can simplify our analysis by choosing a frame in which the eigenvectors of $\mb Q$ and $\mb X$ coincide with the Cartesian basis $(\mb e_x,\mb e_y,\mb e_z)$, where $\mb e_x=(1,0,0), \mb e_y = (0,1,0), \mb e_z = (0,0,1)$. Then $\mb Q$ (the solutions of \eqref{bulkEL}) and $\mb X$ are diagonal in the Cartesian coordinates and can be expressed as
\begin{align}
    \mb Q &= S \left(\mb e_x\otimes\mb e_x-\tfrac{1}{3}\mb I\right) + R \left(\mb e_y\otimes\mb e_y-\tfrac{1}{3}\mb I\right), \label{Q}\\
    \mb X & = s_* \left(\mb e_x\otimes\mb e_x-\tfrac{1}{3}\mb I\right) + r_* \left(\mb e_y\otimes\mb e_y-\tfrac{1}{3}\mb I\right). \label{X}
\end{align}
Here, $S$ and $R$ are two unknown scalar order parameters while $s_*$ and $r_*$ are given scalar order parameters prescribed by $\mb X$ in \eqref{Xtensor}. The critical points of $f_m$ are determined by $(S,R)$, which in turn, are solutions of the following coupled algebraic equations:
\begin{align}
    S\left[\mc A + \Sigma_s W -\tfrac{1}{3}\mc B(S-2R) + \tfrac{2}{3}(S^2-SR+R^2) \right]    &= W s_*, \label{SR1}\\
    R\left[\mc A + \Sigma_s W -\tfrac{1}{3}\mc B(R-2S) + \tfrac{2}{3}(S^2-SR+R^2) \right]    &= W r_*. \label{SR2}
\end{align}
These equations depend on the six parameters, $\mc A$, $\mc B$, $W$, $\Sigma_s$, $s_*$ and $r_*$ which only appear in four distinct combinations. As a result, it is often more convenient to use the following four redefined parameters
\begin{equation}
    \tilde{\mc A}\equiv \mc A+\Sigma_s W, \quad \mc B, \quad \tilde s_*\equiv W s_*, \quad \tilde r_*\equiv W r_*.
\end{equation}

The coupled equations~\eqref{SR1}-\eqref{SR2} can be reduced to a single septic polynomial for either $S$ or $R$, computed  using Maple software and verified by hand. For $S$, the septic polynomial takes the form
\begin{align} \label{septic}
    &4\mc B^2 S^7 -(\mc B^4 - 12 \tilde{\mc A}\mc B^2 + 8 \mc B\tilde r_*-4\mc B \tilde s_*)S^5 - 10\mc B^2\tilde s_* S^4 + \nonumber \\ & [2\mc B^3(2\tilde r_*-\tilde s_*)  + 9 \tilde{\mc A}^2\mc B^2 +6\tilde{\mc A}\mc B(\tilde s_*-2\tilde r_*) + 4(\tilde s_*^2-\tilde s_*\tilde r_*+\tilde r_*^2)] S^3 \nonumber \\ &
    -4\mc B\tilde s_*(3\tilde{\mc A}\mc B- 4 \tilde r_* + 2 \tilde s_*) S^2 + 3\tilde s_*^2(\mc B^2 + 2 \tilde{\mc A}) S - 6\tilde s_*^3 =0.
\end{align}
Once $S$ is determined, $R$ is calculated from the sister relation\footnote{\red{By eliminating the common terms inside the brackets in~\eqref{SR1}-\eqref{SR2}, we obtain the relation $\mc BSR (S-R)=\tilde r_*S-\tilde s_* R$. Substituting this relation in~\eqref{SR1} after rewriting it as $2\mc BS R-2SR(S-R)=-(2S^3 - \mc B S^2 + 3\tilde{\mc A}S - 3 \tilde s_* )$ yields $2\mc B^2 SR-2\tilde r_*S+2\tilde s_*R=-\mc B (2S^3 - \mc B S^2 + 3\tilde{\mc A}S - 3 \tilde s_* )$. The last relation allows to express $R$ in terms of $S$ (equation~\eqref{Requation}). Once $R=R(S)$ is found, we can use~\eqref{SR1} or~\eqref{SR2} to arrive at the septic polynomial.}}
\begin{equation}
    R = \frac{2\tilde r_* S - \mc B(2S^3 - \mc B S^2 + 3\tilde{\mc A}S - 3 \tilde s_* )}{2(\tilde s_* +\mc B^2 S)} \quad \text{provided} \quad S,\tilde s_*\neq 0.  \label{Requation}
\end{equation}
For the special case $S=\tilde s_*=0$, the septic polynomial in $R$ can be used to compute the critical points of $f_m$. In general, this allows for up to seven distinct solutions for the pair $(S,R)$. For reasons that will become clear in the subsequent sections, the seven roots are labelled by $(S_0,R_0)$ and $(S_\pm^i,R_\pm^i)$ with $i=x,y,z$.

In the following sections, we compute the roots $(S,R)$ from \eqref{septic} and classify them as isotropic, uniaxial or biaxial phases. The degree of biaxiality is often measured in terms of the biaxiality parameter, $\beta$, defined to be 
\begin{equation}
    \beta = 1 - 6\frac{(\tr\,\mb Q^3)^2}{(\tr\,\mb Q^2)^3} = \frac{27 S^2 R^2 (S-R)^2}{4(S^2-SR+R^2)^3}, \quad \beta\in[0,1].
\end{equation}
A corresponding biaxiality parameter $\beta_*$  can be similarly defined for $\mb X$ using $(s_*,r_*)$. There is close correspondence between the biaxiality of $\mb X$ and that of the critical points of $f_m$ in \eqref{modified}. Furthermore, based on $\mathrm{det}\,\mb Q =[2(S^3+R^3)-3SR(S+R)]/27$, the critical points are classified as prolate $(\mathrm{det}\,\mb Q>0)$ or oblate $(\mathrm{det}\,\mb Q<0)$ about the axis corresponding to the eigenvalue that has a different sign in the triad of eigenvalues: $(\lambda_1,\lambda_2,\lambda_3)=(2S-R,2R-S,-S-R)/3$. 
The stability of each critical point is assessed by calculating the eigenvalues of the Hessian matrix of the modified bulk potential~\eqref{modified} with respect to $(S,R)$. The two eigenvalues of the Hessian are given by
\begin{equation}
    \frac{1}{2}(f_{m,SS}+f_{m,RR}) \pm \frac{1}{2}\sqrt{(f_{m,SS}-f_{m,RR})^2+4f_{m,SR}}
\end{equation}
where $f_{m,SS}=\tfrac{2}{3} \tilde{\mc A} + \tfrac{2}{3}(2S^2+R^2) - \tfrac{2}{9}\mc B (2S-R) - \tfrac{4}{3}SR$, $f_{m,RR}=\tfrac{2}{3} \tilde{\mc A} + \tfrac{2}{3}(2R^2+S^2) - \tfrac{2}{9}\mc B (2R-S) - \tfrac{4}{3}SR$ and $f_{m,SR}=-\tfrac{1}{3} \tilde{\mc A} - \tfrac{2}{3}(S^2+R^2) + \tfrac{2}{9}\mc B (S+R) + \tfrac{4}{3}SR$.

\subsection{Limiting solutions for small and large values of $W$}
\label{sec:limiting}

The critical points of $f_m$ can be explicitly computed for $W=0$ as detailed in ~\cite{dg,majumdar2010equilibrium,mottram2014introduction}. These critical points correspond to either the isotropic state, $(S_0,R_0)=(0,0)$, or  uniaxial states for $\mc A<\mc B^2/24$. Specifically, there are six uniaxial critical points for $W=0$, given by $(S_\pm^x,R_\pm^x)=(s_\pm,0)$, $(S_\pm^y,R_\pm^y)=(0,s_{\pm})$ and $(S_\pm^z,R_\pm^z)=(-s_\pm,-s_{\pm})$, where $s_\pm = \mc B/4 \pm \sqrt{\mc B^2-24\mc A}/4$. Each pair of critical points represents uniaxial states along one of the three Cartesian axes. Furthermore, when $\mc A<0$, all three $s_+$ states are prolate and the $s_-$ states are oblate. The global minimizer of $f_m$ (with $W=0$) is the isotropic state when $\mc A>\mc B^2/27$ and it pertains to any one of the three $s_+$ states when $\mc A<\mc B^2/27$. The uniaxial states parameterized by $s_-$ are locally stable for $\mc A <0$ and unstable for $\mc A>0$.

Consider small non-zero values of $W$ with $\mb X \neq 0$, so that we can compute an asymptotic expansion of the seven critical points of $f_m$, as a polynomial in $W$ (using the same notation for the critical points as for $W=0$). Keeping $\mc A$ and other parameters fixed, with errors of order $W^2$, 
\begin{align}
    (S_0,R_0) &= (0,0) +  \frac{W}{\mc A}(s_*,r_*), \nonumber \\
      (S_\pm^x,R_\pm^x) &= (s_\pm,0)  + \frac{W}{\mc B s_\pm}\left(\frac{\mc B [3s_\pm \Sigma_s-3s_*+2r_*] - 2r_*s_\pm }{ \mc B-4s_\pm},r_*\right) , \nonumber \\
     (S_\pm^y,R_\pm^y) &= (0,s_\pm)  + \frac{W}{\mc B s_\pm}\left(s_*,\frac{\mc B [3s_\pm \Sigma_s+2s_*-3r_*] - 2s_*s_\pm }{ \mc B-4s_\pm}\right),  \nonumber \\
      (S_\pm^z,R_\pm^z) &= (-s_\pm,-s_\pm) + \frac{W}{\mc B s_\pm}\left( \frac{\mc B [3s_\pm \Sigma_s+s_*+2r_*] + 2s_\pm (s_*-r_*)}{\mc B-4s_\pm},\right.  \nonumber \\ 
 &\qquad \qquad \qquad \qquad \qquad \,\, \left. \frac{\mc B [3s_\pm \Sigma_s+2s_*+r_*] - 2s_\pm (s_*-r_*)}{\mc B-4s_\pm}\right)\nonumber .
\end{align}
The biaxiality parameter of the continuation of the isotropic branch, denoted by $(S_0,R_0)$ is $\beta_0=\beta_*+O(W)$. In the remaining six cases of the continuation of the uniaxial critical points, the biaxiality parameter is small and is given by $\beta_\pm^x=27 W^2 r_*^2/4\mc B^2 s_\pm^4$, $\beta_\pm^y=27 W^2 s_*^2/4\mc B^2 s_\pm^4$ and $\beta_\pm^z=27 W^2 (s_*-r_*)^2/4\mc B^2 s_\pm^4$, upon neglecting higher-order corrections. It is useful to recognize that the above asymptotic expansions are valid not only for $W\to 0$, but also for $|\mc A|\to\infty,\mc A<0$. This is because $1/s_{\pm} \to 0$ as $|\mc A|\to\infty,\mc A<0$ and consequently the small-$W$ expansion also extends to the low-temperature expansion. For example, the first correction to the solution $(s_\pm,0)$ as $\mc A\to-\infty$ is given by
\begin{equation}
    (S_\pm^x,R_\pm^x) = (s_\pm,0) \pm \frac{W (2r_*-3\mc B\Sigma_s,4r_*) }{\mc B\sqrt{24|\mc A|}}.
\end{equation}

Turning to the limit $W\to \infty$ for fixed $\mc A$, the modified bulk potential in~\eqref{modified} simplifies, as a first approximation, to $CW[\tfrac{1}{2}\Sigma_s\tr\,\mb Q^2 - \tr(\mb Q\mb X)]$, which is the net homogenized NP-LC interaction energy density. This leading order modified bulk energy is minimized uniquely by $\mb Q = \mb X/\Sigma_s$ or $(S,R)=(s_*/\Sigma_s,r_*/\Sigma_s)$. In this limiting case, the biaxiality parameter of the energy minimizer, $\mb Q$, is equal to that of $\mb X$, i.e., $\beta=\beta_*$. On the other hand, as $\mc A \to +\infty$ for a fixed $W$, $f_m$ is uniquely minimized by an almost isotropic phase, i.e., $\mb Q\to W\mb X/\mc A$. In other words, the qualitative features of the original bulk potential, $f_B$ in \eqref{eq:2}, are preserved in the $\mc A \to \pm \infty$ limits, for fixed $W$.

\subsection{Critical points for isotropic $\mb X = 0$}

When the tensor $\mb X$ is isotropic, i.e., $\mb X=0$, the collective effect of the NPs is also isotropic and the effective temperature is translated from $\mc A$ to $\tilde{\mc A}=\mc A +\Sigma_s W$. In other words, the effective temperature increases in the presence of colloidal NPs when $ W >0$. The critical points are then the same as the seven solutions mentioned at the beginning of  Section~\ref{sec:limiting} for $W=0$, except that the isotropic phase is stable for $\tilde{\mc A}>0$ or for $\mc A > - \Sigma_s W$; the isotropic-nematic transition occurs at $\tilde{\mc A}=\mc B^2/27$ or for $\mc A = \mc B^2/27 - \Sigma_s W$ and the ordered LC states do not exist for $\mc A > \mc B^2/24 - \Sigma_s W$. Thus, the colloidal NPs broaden the window of stability of the isotropic phase and shrink the domain of existence for the ordered LC states.

 Note that $\mb Q=0$ is not a critical point of $f_m$ for $\mb X \neq 0$, as can be seen from~\eqref{bulkEL}. 

\subsection{Critical points for uniaxial $\mb X$}

When $\mb X$ is uniaxial, either prolate or oblate, the septic polynomial~\eqref{septic} can be factored into a cubic and a quartic polynomial, thus simplifying the analysis. Without loss of generality, we assume that 
\begin{equation}
    \mb X = s_* \left(\mb e_x\otimes\mb e_x -\tfrac{1}{3}\mb I\right),  \label{Xuni}
\end{equation}
where $\mb e_x=(1,0,0)$, i.e., $r_*=0$ in~\eqref{X}. The uniaxial $\mb X$ is prolate if $s_*>0$ and oblate if $s_*<0$.

\subsubsection{Prolate uniaxial $\mb X$}

We first consider the case of a prolate uniaxial $\mb X$ tensor, assuming $s_*>0$. There are seven critical points again, three of which are uniaxial along the $x$-axis or the director of $\mb X$, and four of which are biaxial with the leading director along the $y$- and $z$-axes respectively. These seven critical points can be viewed as a continuation of the seven critical points of $f_m$ with $W=0$. The critical points, $(S,R) = (0,0)$ and $(S,R)= (s_\pm, 0)$ (for $W=0$), are continued to uniaxial critical points along the $x$-axis for prolate $\mb X$ as in \eqref{Xuni}, when $W>0$. They are denoted by $(S_0,0)$ and $(S_{\pm}^x, 0)$ respectively. On the other hand, the critical points, $(S,R) = (0, s_\pm)$ and $(S,R) = (-s_\pm, -s_\pm)$ (for $W=0$) get continued to biaxial critical points, denoted by $(S_{\pm}^y, R_{\pm}^y)$ and $(S_{\pm}^z, R_{\pm}^z)$ respectively, for $W>0$.   A summary of these seven critical points and their stability properties is given in Table~\ref{tab:prolateX}.

\begin{table}[htbp]
\footnotesize
\caption{The critical points of $f_m$ in \eqref{modified} for prolate uniaxial $\mb X$: $\mb X= s_* \left(\mb e_x\otimes\mb e_x -\tfrac{1}{3}\mb I\right),\,s_*>0$. Here, $\tilde{\mc A}_4^+$ defines the existence range for $(S_+^{y,z},R_+^{y,z})$, determined by the quartic polynomial~\eqref{quartic}; see, Figure~\ref{fig:prolateX}(b).}\label{tab:prolateX}
  \begin{tabular}{|c|c|} \hline
    Solution type & Stability \\ \hline
     $(S_0,0)$, uniaxial, oblate ($x$-axis) & unstable \\
    $(S_+^x,0)$, uniaxial, prolate ($x$-axis) & global minimum ($\forall \tilde{\mc A}$) \\ 
     $(S_-^x,0)$, uniaxial, oblate ($x$-axis) & unstable \\ 
     $(S_+^y,R_+^y)$, biaxial, prolate ($y$-axis) & local minimum ($\tilde{\mc A}<\tilde{\mc A}_4^+$)  \\ 
     $(S_-^y,R_-^y)$, biaxial, oblate ($y$-axis) & unstable \\ 
    $(S_+^z,R_+^z)$, biaxial, prolate ($z$-axis) & local minimum ($\tilde{\mc A}<\tilde{\mc A}_4^+$) \\ 
    $(S_-^z,R_-^z)$, biaxial, oblate ($z$-axis) & unstable \\ \hline
  \end{tabular}
\end{table} 

\begin{figure}[h!]
\centering
\includegraphics[width=0.45\textwidth]{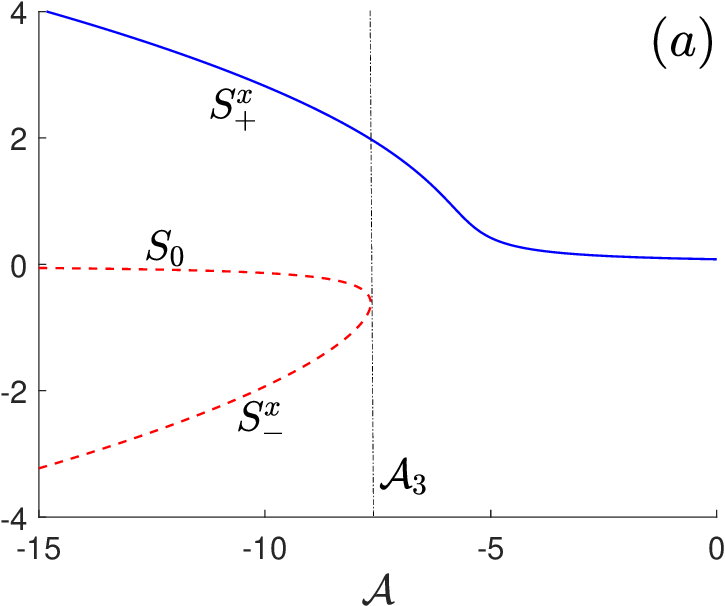}
\includegraphics[width=0.45\textwidth]{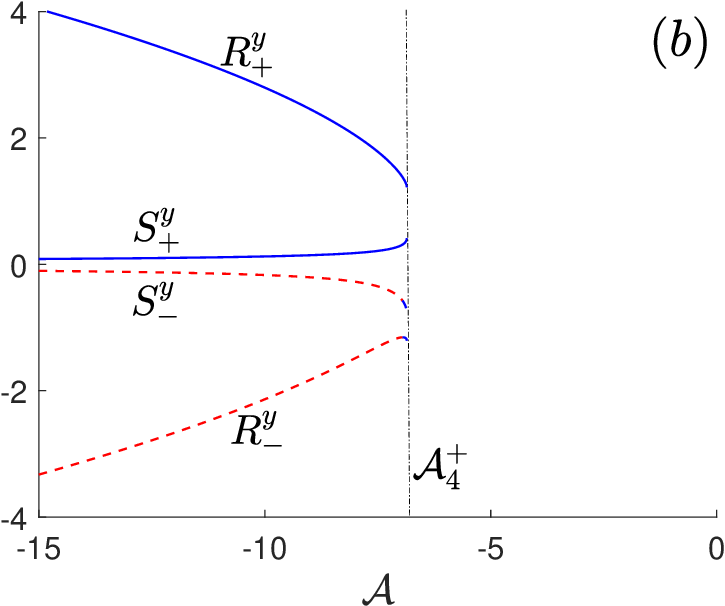}\vspace{0.5cm}
\includegraphics[width=0.45\textwidth]{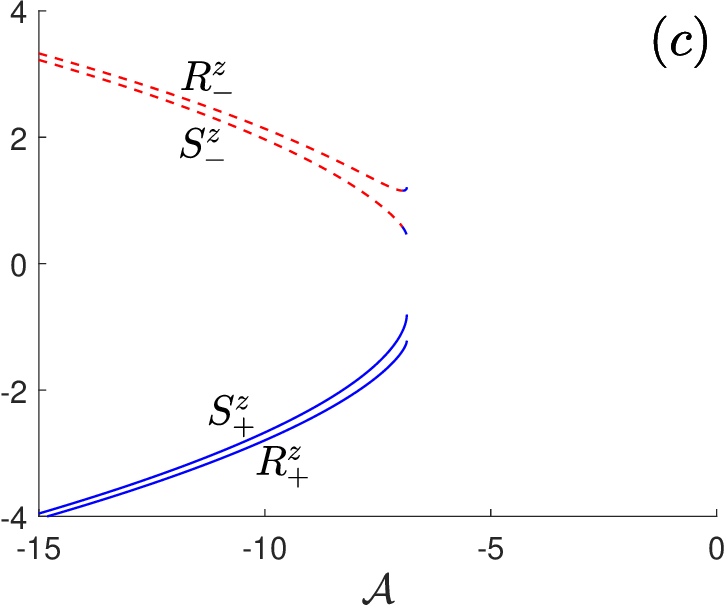}
\includegraphics[width=0.45\textwidth]{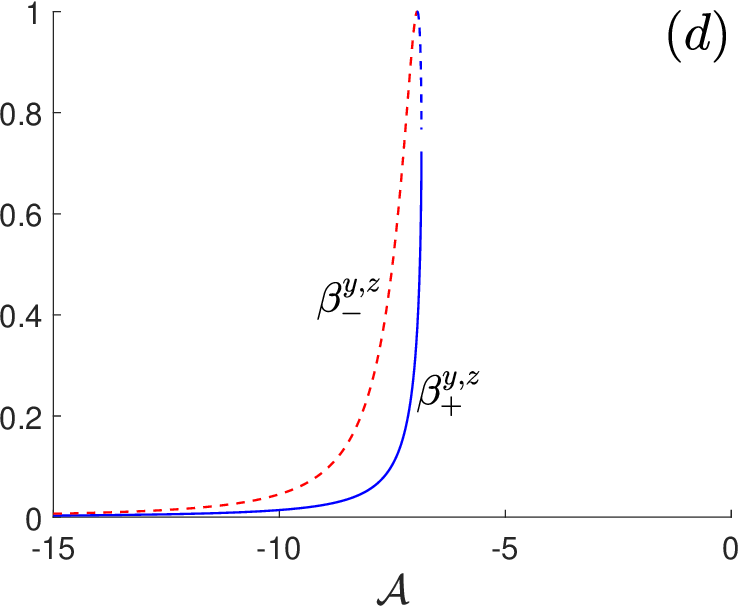}
\caption{Critical points of $f_m$ for prolate uniaxial $\mb X$ (about $x$-axis) with $\mc B=1.5$, $\Sigma_s=4\pi$, $W=0.5$ and $(s_*,r_*)=(1,0)$. Recall that $\mc A = \tilde{\mc A} - \Sigma_s W$. In all plots, solid lines indicate stable critical points and dashed lines indicate unstable critical points. Similarly, blue lines indicate prolate symmetry and red ones oblate symmetry. (a) Three uniaxial solutions about $x$-axis, which as $\mc A\to-\infty$ approach the two uniaxial solutions about $x$-axis and the isotropic branch. (b) Biaxial solutions, which as $\mc A\to-\infty$ approach the uniaxial solutions about $y$-axis. (c) Biaxial solutions, which as $\mc A\to-\infty$ approach the uniaxial solutions about $z$-axis. (d) Biaxiality parameter $\beta$ pertaining to plot (b) and (c).} 
\label{fig:prolateX}
\end{figure}

The septic polynomial in \eqref{septic} can be factored into a cubic polynomial and a quartic polynomial, for a uniaxial $\mb X$. The cubic polynomial for $S$ determines the uniaxial critical points of $f_m$ in \eqref{modified} and is factored from~\eqref{septic} as shown below:
\begin{align}
     2S^3 - \mc B S^2 + 3\tilde{\mc A} S - 3 \tilde s_*=0, \qquad R=0. \label{cubic}
\end{align}
The three roots of this polynomial correspond to the first three uniaxial solutions, $S_0$ and $S_\pm^x$, listed in Table~\ref{tab:prolateX}. The uniaxial solution, $(S_0,0)$, which emerges from the isotropic branch, only exists when $\tilde{\mc A} < \tilde{\mc A}_3 $ and is oblate when $\mb X$ is prolate, and vice versa. The discriminant of the cubic polynomial, $\Delta = 9\tilde{\mc A}^2\mc B^2 - 12 \tilde s_* \mc B^3 - 216 \tilde{\mc A}^3 + 324 \tilde s_* \tilde{\mc A}\mc B - 972 \tilde s_*^2$, shows that there are three roots for sufficiently large and negative values of $\tilde{\mc A}$ (when $\Delta>0$). Two of the roots coalesce at a critical value $\tilde{\mc A}_3$, where $\Delta=0$, and there exists only one real root for $\tilde{\mc A}>\tilde{\mc A}_3$ or for $\mc A > \tilde{\mc A}_3 -\Sigma_s W$.
An illustrative plot of the three roots as a function of temperature, $\mc A = \tilde{\mc A} - \Sigma_s W$ is shown in Figure~\ref{fig:prolateX}(a).

On the other hand, the quartic polynomial for $S$ and the corresponding $R$ values are given by
\begin{align}
    2\mc B^2 S^4 + \mc B^3 S^3 + \mc B(3\tilde{\mc A}\mc B+2\tilde s_*) S^2 - \mc B^2 \tilde s_* S + 2\tilde s_*^2 =0,  \label{quartic}\\
     R = \frac{ - \mc B(2S^3 - \mc B S^2 + 3\tilde{\mc A}S - 3 \tilde s_* )}{2(\tilde s_* +\mc B^2 S)}. \label{quarticR}
\end{align}
The roots of the quartic polynomial represent the remaining four solutions from Table~\ref{tab:prolateX}. By examining the quartic polynomial, we can determine the nature of these four roots as $|\tilde{\mc A}|\to \infty$. As $\tilde{\mc A}\to-\infty$, all four roots are real and distinct, approaching the four uniaxial states (labelled by $(S,R)=(0, s_{\pm})$ and $(S,R)=(-s_{\pm}, -s_{\pm})$ respectively) along the $y$- and $z$-axes. Conversely, as $\tilde{\mc A}\to+\infty$, none of the roots are real, and thus no solutions exist. These four roots are plotted in Figure~\ref{fig:prolateX}(b) and Figure~\ref{fig:prolateX}(c). As expected, the level of biaxiality between the two axes is the same, i.e., $\beta_+^y=\beta_+^z$ and $\beta_-^y=\beta_-^z$. Figure~\ref{fig:prolateX}(d) illustrates these biaxiality parameters graphically, complementing Figure~\ref{fig:prolateX}(b) and Figure~\ref{fig:prolateX}(c).

In summary, there are seven distinct critical points of $f_m$, for low temperatures, including co-existing locally stable biaxial and globally stable (global energy minimizers) uniaxial bulk phases. The locally stable biaxial phases are induced by the colloidal NPs and cannot be observed with $W=0$ or in the pure cholesteric system. The two biaxial local energy minimizers, $(S,R)=(S_+^y,R_+^y)$ and $(S,R)=(S_+^z,R_+^z)$,  exist only for $\tilde{\mc A} < \tilde{\mc A}_4^+$, as demarcated in Figure~\ref{fig:prolateX}. There is no isotropic-nematic/cholesteric phase transition as such, when $W \neq 0$ and $\mb X \neq 0$. At high temperatures, we only observe the globally stable uniaxial critical point, $(S,R) = (S_+^x, 0)$, where $S_+^x \to \tilde{s_*}/\mc A$ as $A \to + \infty$. Furthermore, the critical point, $(S,R)=(S_+^x,0)$, is the global minimizer of the modified bulk potential for all temperatures.  

\begin{table}[htbp]
\footnotesize
\caption{The critical points of $f_m$ in \eqref{modified} for oblate uniaxial $\mb X$: $\mb X= s_* \left(\mb e_x\otimes\mb e_x -\tfrac{1}{3}\mb I\right),\,s_*<0$. Here, $\tilde{\mc A}_4^+$ defines the existence range for $(S_+^{y,z},R_+^{y,z})$, determined by the quartic polynomial~\eqref{quartic} and $\tilde{\mc A}_3$ 
 defines the existence range for $(S_+^x,0)$, determined by the cubic polynomial~\eqref{cubic}.}\label{tab:oblateX}
  \begin{tabular}{|c|c|} \hline
   Solution type & Stability  \\ \hline
    $(S_0,0)$, uniaxial, prolate ($x$-axis) & unstable \\
    $(S_+^x,0)$, uniaxial, prolate ($x$-axis) &  local minimum ($\tilde{\mc A}<\tilde{\mc A}_3$) \\ 
     $(S_-^x,0)$, uniaxial, oblate ($x$-axis) &  global minimum  ($\tilde{\mc A}>\tilde{\mc A}_4^+$) \\ 
     $(S_+^y,R_+^y)$, biaxial, prolate ($y$-axis) &   global minimum  ($\tilde{\mc A}<\tilde{\mc A}_4^+$) \\ 
     $(S_-^y,R_-^y)$, biaxial, oblate ($y$-axis) & unstable \\ 
     $(S_+^z,R_+^z)$, biaxial, prolate ($z$-axis) & global minimum  ($\tilde{\mc A}<\tilde{\mc A}_4^+$) \\ 
     $(S_-^z,R_-^z)$, biaxial, oblate ($z$-axis) & unstable \\ \hline
  \end{tabular}
\end{table} 

\begin{figure}[h!]
\centering
\includegraphics[width=0.45\textwidth]{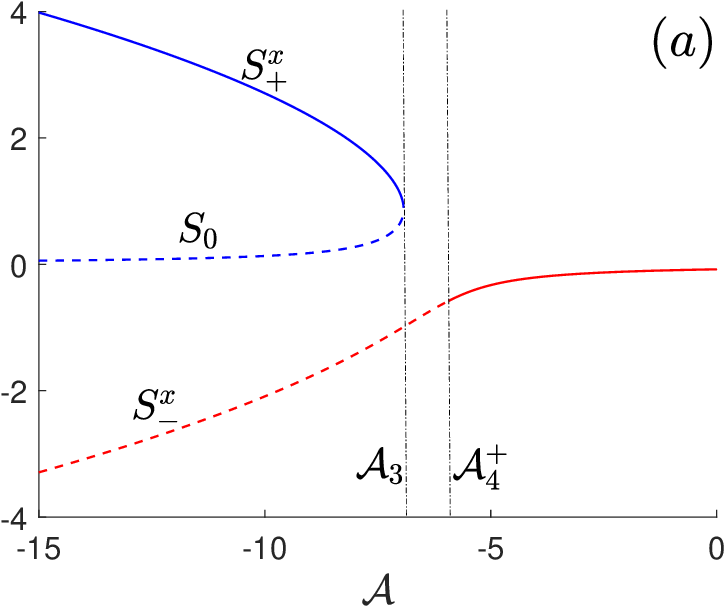}
\includegraphics[width=0.45\textwidth]{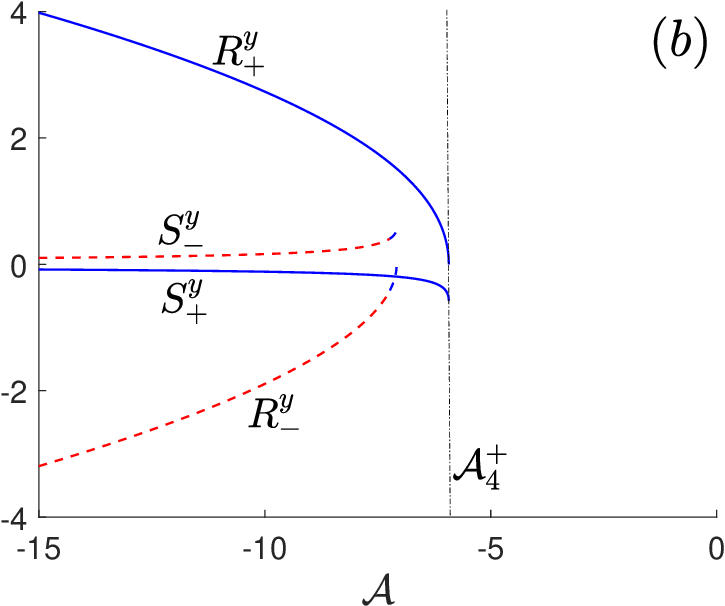}\vspace{0.5cm}
\includegraphics[width=0.45\textwidth]{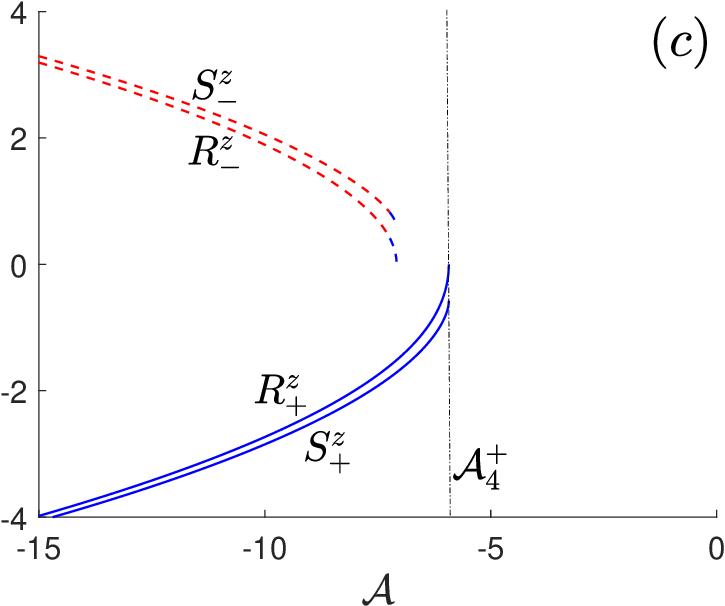}
\includegraphics[width=0.45\textwidth]{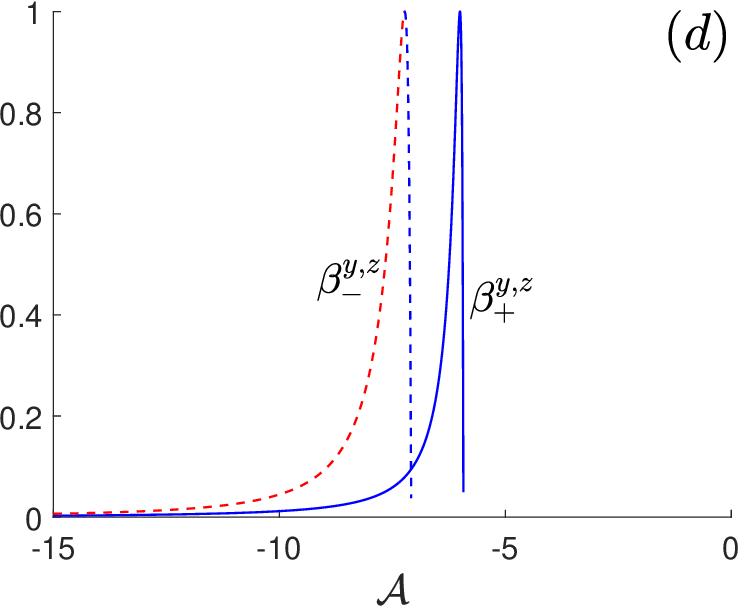}
\caption{Critical points of $f_m$ for oblate uniaxial $\mb X$ (about $x$-axis) with $\mc B=1.5$, $\Sigma_s=4\pi$, $W=0.5$ and $(s_*,r_*)=(-1,0)$. In all plots, solid lines indicate stable critical points and dashed lines indicate unstable critical points. Similarly, blue lines indicate prolate symmetry and the red ones oblate symmetry. (a) Three uniaxial solutions about the $x$-axis. (b) Two biaxial solutions about the $y$-axis. (c) Two biaxial solutions about the $z$-axis. (d) Biaxiality parameter $\beta$ pertaining to plot (b) and (c).} 
\label{fig:oblateX}
\end{figure}

\subsubsection{Oblate uniaxial $\mb X$}

When $s_*$ is negative in~\eqref{Xuni}, the prescribed $\mb X$ tensor exhibits oblate symmetry about the $x$-axis. While the same cubic and quartic equations~\eqref{cubic}-\eqref{quarticR} still apply, the stability properties are significantly different from the prolate case. We still have three uniaxial critical points from the cubic polynomial: $(S_0,0)$, $(S_+^x,0)$, $(S_-^x, 0)$ such that $S_+^x$ is locally stable for $\mc A < \mc A_3$ and does not exist for $\mc A > \mc A_3$. This is in stark contrast to the case of prolate uniaxial $\mb X$, for which $(S_+^x, 0)$ is globally stable for all $\mc A$. The critical point, $(S, R) = (S_0, 0)$, which is the continuation of the isotropic branch, has prolate symmetry for oblate $\mb X$. The oblate uniaxial critical point, $(S_-^x, 0)$, is unstable for $\mc A < \mc A_4^+$ but becomes the unique globally stable critical point for high temperatures, i.e., for $\mc A > \mc A_4^+$.  Therefore, there is an exchange of stability between prolate and oblate critical points at  $\mc A=\mc A_4^+$. The globally stable prolate biaxial critical points, $(S_+^y, R_+^y)$ and $(S_+^z, R_+^z)$  exist only for $\mc A\leq \mc A_4^+$. The oblate biaxial critical points are always unstable. Notably, there are only stable biaxial phases in a narrow temperature range, $\mc A_3 < \mc A < \mc A_4^+$, which is a novelty induced by the suspended NPs. As with the case of uniaxial prolate $\mb X$, the oblate uniaxial critical point approaches $S_-^x\to \tilde s_*/\mc A<0$ as $\mc A \to +\infty$ and there is no isotropic-nematic/cholesteric phase transition as such. However, there is co-existence between globally stable biaxial phases and the locally stable prolate uniaxial phase for $\mc A < \mc A_3$. The seven critical points are plotted in Figure~\ref{fig:oblateX} and their stability properties are summarised in Table~\ref{tab:oblateX}. 
 
\begin{figure}[h!]
\centering
\includegraphics[width=0.45\textwidth]{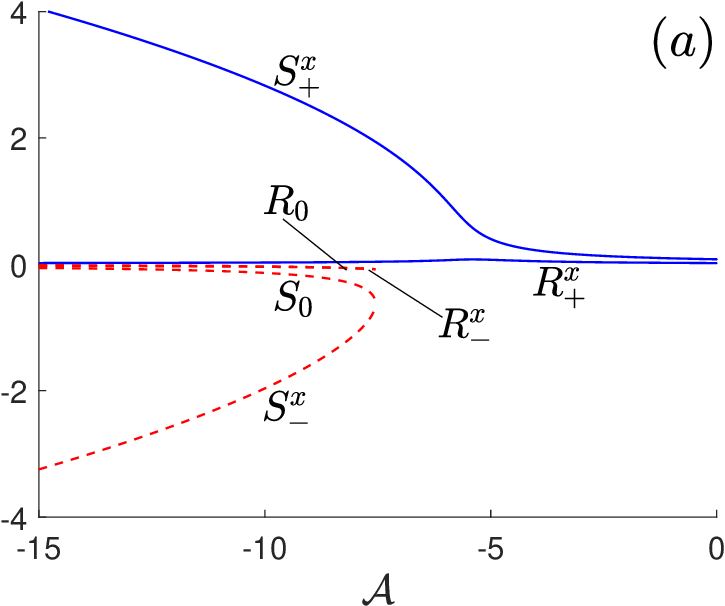}
\includegraphics[width=0.45\textwidth]{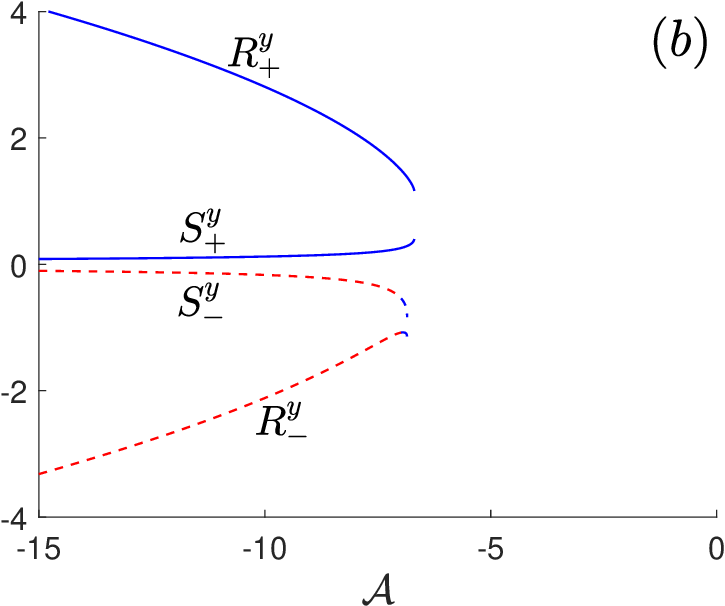}\vspace{0.5cm}
\includegraphics[width=0.45\textwidth]{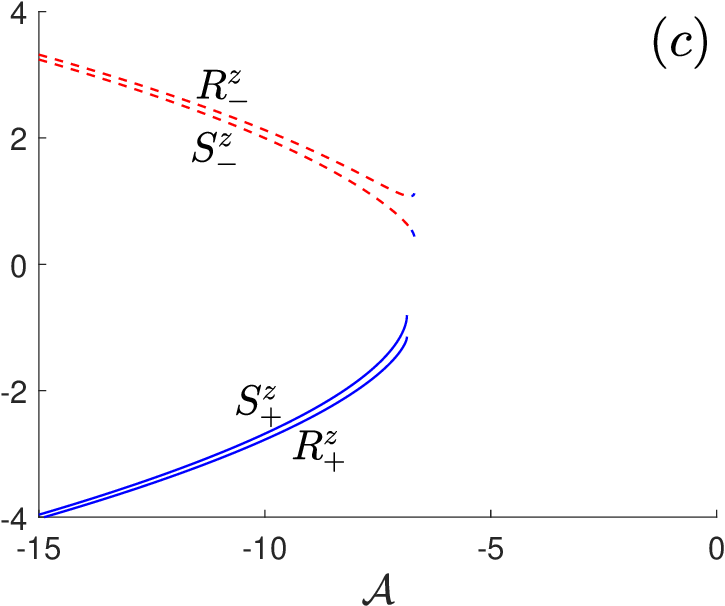}
\includegraphics[width=0.45\textwidth]{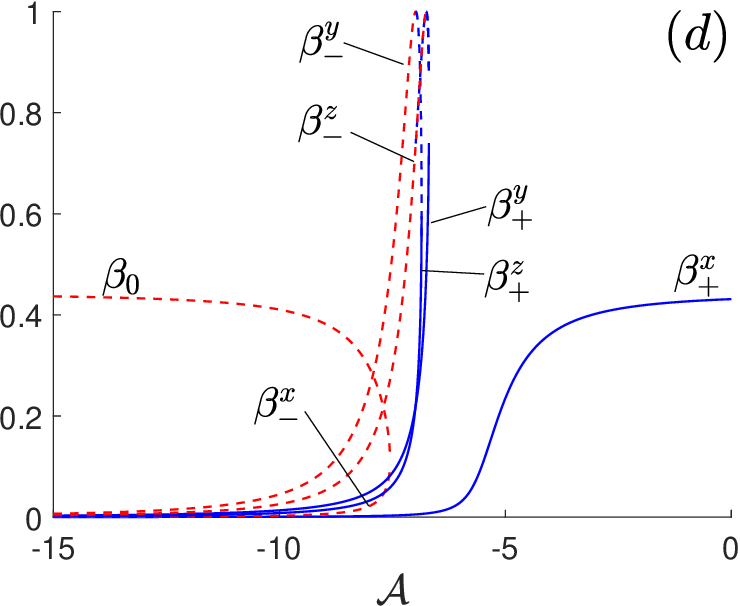}
\caption{The critical points of $f_m$ for prolate biaxial $\mb X$ (about $x$-axis) with $\mc B=1.5$, $\Sigma_s=4\pi$, $W=0.5$ and $(s_*,r_*)=(1,1/4)$; the biaxiality parameter of $\mb X$ is $\beta_*=0.44$. In all plots, solid lines indicate stable critical points and dashed lines indicate unstable critical points. Similarly, blue lines indicate prolate symmetry and the red ones oblate symmetry. (a) Three biaxial solutions about the $x$-axis. (b) Two biaxial solutions about the $y$-axis. (c) Two biaxial solutions about the $z$-axis. (d) The biaxiality parameter $\beta$ of the seven solutions.} 
\label{fig:biaxial1}
\end{figure}

\begin{figure}[h!]
\centering
\includegraphics[width=0.45\textwidth]{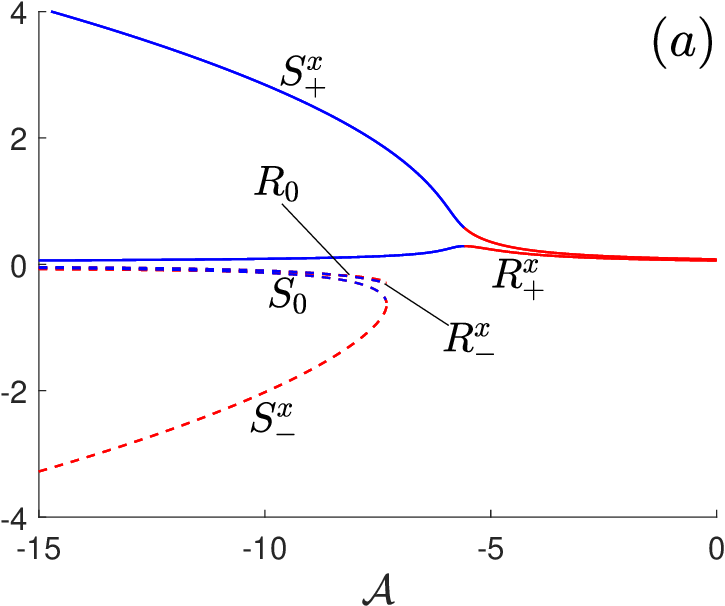}
\includegraphics[width=0.45\textwidth]{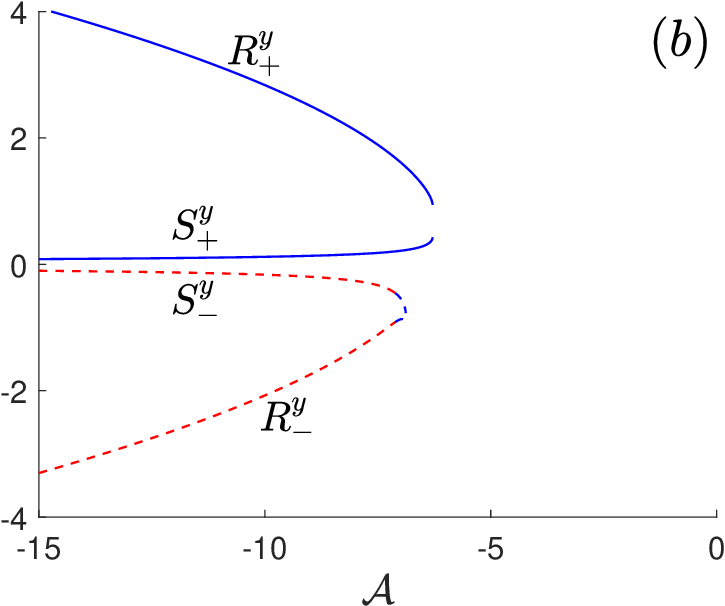}\vspace{0.5cm}
\includegraphics[width=0.45\textwidth]{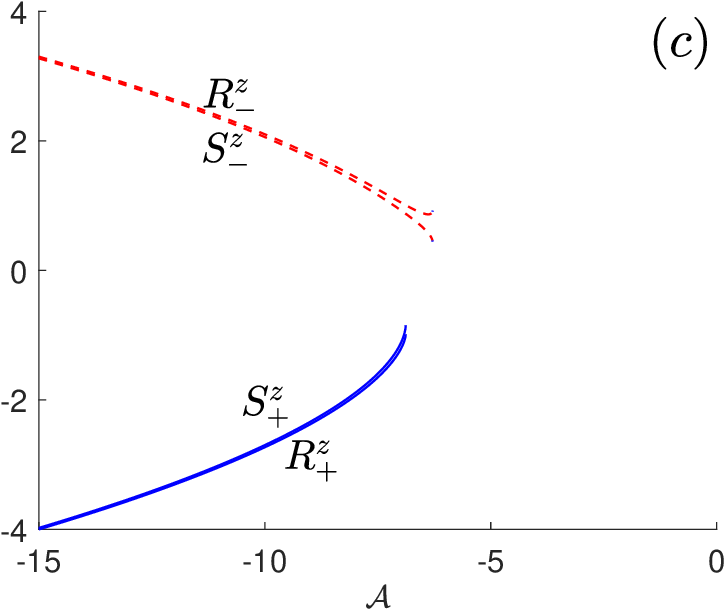}
\includegraphics[width=0.45\textwidth]{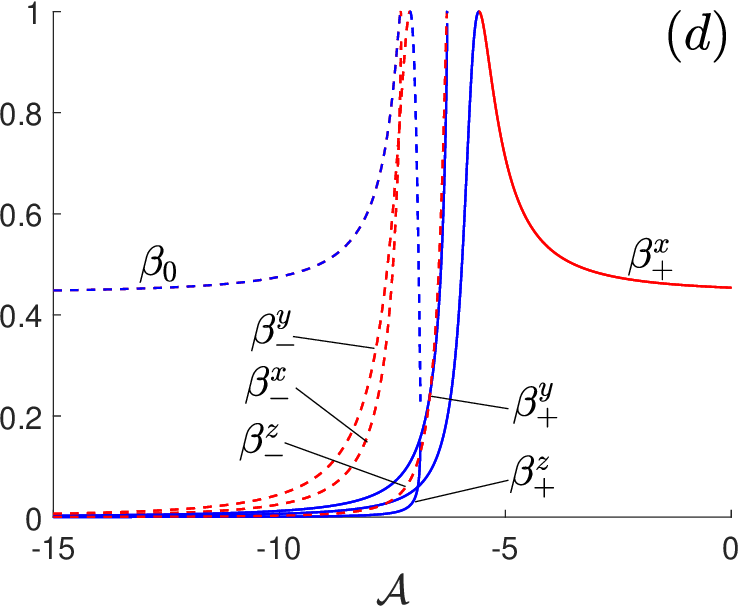}
\caption{The critical points of $f_m$ for oblate biaxial $\mb X$ (about $z$-axis) with $\mc B=1.5$, $\Sigma_s=4\pi$, $W=0.5$ and $(s_*,r_*)=(1,3/4)$; the biaxiality parameter of $\mb X$ is $\beta_*=0.44$. In all plots, solid lines indicate stable critical points and dashed lines indicate unstable critical points. Similarly, blue lines indicate prolate symmetry and the red ones oblate symmetry. (a) Three biaxial solutions about the $x$-axis. (b) Two biaxial solutions about the $y$-axis. (c) Two biaxial solutions about the $z$-axis. (d) The biaxiality parameter $\beta$ of the seven critical points.} 
\label{fig:biaxial2}
\end{figure}

\subsection{Critical points for biaxial $\mb X$}

Next, we compute the critical points of $f_m$ in \eqref{modified} with a prescribed biaxial $\mb X$. Given the complexity of classifying the solution landscape in the general biaxial case, we focus only on two specific examples defined by $(s_*,r_*)=(1,1/4)$ and $(s_*,r_*)=(1,3/4)$ respectively. Both cases have the biaxiality parameter, $\beta_*=0.44$. In the first case, $\det \mb X >0 $ and the eigenvalues of $\mb X$ are  $(7,-2,-5)/12$. Therefore, $\mb X$ is a prolate biaxial state about the $x$-axis, with a preference for alignment along the $y$-axis over the $z$-axis. In the second case, $\det \mb X <0 $ and the eigenvalues of $\mb X$ are $(5,2,-7)/12$. Thus, $(s_*,r_*)=(1,3/4)$ is an oblate biaxial state about the $z$-axis, with a preference for alignment along the $x$-axis over the $y$-axis. As demonstrated in previous examples, the symmetry of $\mb X$ dictates the symmetry of the critical points as $\mc A \to \infty$.

In both cases, there are seven biaxial critical points as plotted in Figure~\ref{fig:biaxial1} and \ref{fig:biaxial2} respectively. In both cases, the global minimizer is $(S_+^x,R_+^x)$ while the local minimizers are $(S_+^{y,z},R_+^{y,z})$. The local minimizers do not exist for $\mc A > \mc A^*$, for some critical value $\mc A^*$.  Interestingly, while $(S_+^x,R_+^x)$ remains prolate (about the $x$-axis) in the first case of prolate biaxial $\mb X$, it converts to an oblate state (about the $z$-axis) at a finite temperature in the second case, mirroring the symmetry of $\mb X$. Another notable feature is that, while $\beta_+^x$ is monotonic in the first case, it reaches a maximum in the second case, coinciding with the change in symmetry of $S_+^x$. In both cases, the global minimizer $(S_+^x,R_+^x)$ approaches the isotropic state as $\mc A \to \infty$. For example, in Figure~\ref{fig:biaxial1}, $\beta_+^x$ remains very small for low temperatures and then increases to noticeable values for $\mc A > -5$. However, $(S_+^x, R_+^x)$ is biaxial and almost isotropic (with both $S_+^x$ and $R_+^x$ being very small) for $\mc A >-5$. Hence, this is unlikely to lead to macroscopically observable biaxiality for high temperatures. Similarly, the biaxiality of $(S_+^{y,z},R_+^{y,z})$ tend to be (very) small for most of the temperature range of their existence and are unlikely to manifest in macroscopically observable biaxiality. Similar comments apply to Figure~\ref{fig:biaxial2}(a), i.e., the critical points about the $x$-axis (the symmetry axis of $\mb X$) are either approximately uniaxial as $\mc A \to -\infty$ or approximately isotropic (for $\mc A>-5$).  We speculate that the locally stable biaxial critical points, $(S_+^y, R_+^y)$ and $(S_+^z, R_+^z)$ are the best candidates for experimentally observable biaxiality, in a narrow range of low temperatures.

To summarise, there are two main conclusions. At low temperatures ($\mc A\to-\infty$), there are three biaxial minima corresponding to $S_+^i$ ($i=x,y,z$) with a slight preference for the axis corresponding to the largest positive eigenvalue of $\mb X$. The low-temperature energy minima are always prolate, since they inherit the symmetry properties of the minimizers of the original bulk energy, $f_B$ in \eqref{eq:2}. At high temperatures, there is only one critical point, which is also the global energy minimum, corresponding to an almost isotropic but biaxial phase. When $\mb X$ is prolate, the high-temperature energy minimizer of $f_m$ is prolate and the leading director of the energy minimizer coincides with the leading director of $\mb X$. When $\mb X$ is oblate, the high temperature energy minimizer inherits the oblate symmetry of $\mb X$, i.e., it is oblate about the same axis as $\mb X$ and exhibits the same leading director (with the largest positive eigenvalue) as $\mb X$. Hence, whilst the homogenized bulk energy minimizer is always prolate for low temperatures, it can be prolate or oblate depending on the symmetry of $\mb X$ for high temperatures. 

In all cases, a non-zero $\mb X$ (uniaxial or biaxial) leads to locally stable biaxial phases for some low temperatures and the possibility of oblate energy minimizers for oblate $\mb X$, both of which are not allowed with either $W=0$ or $\mb X=0$. \red{Hence, the NPs can introduce new physics into spatially homogeneous nematic or cholesteric systems and open new routes for macroscopically observable biaxiality, which is often an elusive concept.}

\section{Spatially inhomogeneous systems: Cholesteric-filled channel geometry}
\label{sec:inhomogeneous}

In Section~\ref{sec:homogeneous}, we  focus on the bulk equilibrium phases determined by the minimizers and critical points of $f_m$ in \eqref{modified}. The modified bulk potential, $f_m$, does not distinguish between nematic and cholesteric liquid crystal phases \cite{han2022uniaxial}. The suspended NPs essentially break the symmetry of the nematic/cholesteric phase if $\mb X \neq 0$ and impose a preferred direction (or director) on homogeneous samples, for large $W$ or large $\mc A$. 

The next step is to extend this analysis to spatially inhomogeneous samples, such as a dilute suspension of colloidal spherical NPs in a cholesteric-filled channel geometry described in \eqref{eq:omega}. Given that $H\ll L\sim D$, it is reasonable to assume that the structural properties are translationally invariant in $x$ and $y$-directions. Therefore, it suffices to study the cholesteric configurations in a reduced one-dimensional domain:
\[
\widetilde{\Omega}=\left\{ 0 \leq z \leq H \right\}.
\]
We impose fixed Dirichlet boundary conditions on the surfaces $z=0$ and $z=H$:
\begin{equation}
    \mb Q(0)=\mb Q(H) =\mb Q_0 \equiv \begin{bmatrix}
        \tfrac{2}{3}s_+ & 0 & 0 \\
        0 & -\tfrac{1}{3}s_+ & 0 \\
        0 & 0 & -\tfrac{1}{3}s_+
    \end{bmatrix}  \label{planar}
\end{equation}
which corresponds to a uniaxial boundary condition with the leading director aligned along the $x$-axis. Here $s_+$ is prescribed as in \eqref{eq:s+} for a given temperature. We could work with a different order parameter for $\mb Q_0$, but such choices would only lead to quantitative changes in the results.

The equilibrium/physically observable cholesteric configurations are modelled by minimizers of the cholesteric free energy, subject to the imposed boundary conditions in \eqref{planar}. We take the effective cholesteric free energy to be the sum of the elastic energy and the modified bulk energy, which captures the effects of the suspended NPs. This is a formal approximation wherein we formally replace the collective effects of the suspended NPs by the modified bulk energy without any rigorous convergence analysis, in the dilute limit. The effective free energy per unit area is given by ~\cite{wright1989crystalline,fukuda2010cholesteric,han2022uniaxial}
\begin{equation}\label{cholesteric_energy1}
    \mc F[\mb Q] = \int_0^H \left[\frac{L_2}{2}(\nabla\cdot\mb Q)^2+\frac{L_1}{2}|\nabla\times\mb Q + 2 q_0\mb Q|^2 + f_b(\mb Q) + f_{hom}(\mb Q)\right]dz,
\end{equation}
where $L_1$ and $L_2$ are material-dependent elastic constants ~\cite{de1971short,stephen1974physics,alexander2006stabilizing} and $2\pi/q_0$ is the preferred cholesteric pitch. The derivatives of $\mb Q$ are defined by $(\nabla\cdot \mb Q)_i = \partial_j Q_{ij} $ and  $(\nabla\times\mb Q)_{ij}=\ep_{ikl}\partial_kQ_{lj}$; the functions $f_b(\mb Q)$ and $f_{hom}(\mb Q)$ are defined in Section~\ref{sec:homo}. The first two terms in \eqref{cholesteric_energy1} are the elastic energy density in \eqref{eq;ldgenergy} and if $q_0 = 0$, then \eqref{cholesteric_energy1} reduces to the LdG energy for nematics. Indeed, the second (twisting) term distinguishes cholesteric phases from nematic phases.

We rescale $z$ with $H$, $\mc F$ with $L_2/H$ and introduce the non-dimensional parameters $\eta = L_1/L_2$ (elastic anisotropy coefficient), $\sigma = 2q_0 H$ (inverse of scaled pitch) and $\lambda = H^2 C/L_2$ (measure of confinement), along with those defined at the end of Section~\ref{sec:homo}. The non-dimensional cholesteric free energy can then be written as
\begin{align}\label{cholesteric_energy}
    \mc F[\mb Q] = \int_0^1 &\left\{\frac{1}{2}(\nabla\cdot\mb Q)^2+\frac{\eta}{2}|\nabla\times\mb Q +  \sigma \mb Q|^2 + \lambda\left[ \frac{\mc A+\Sigma_s W}{2} \tr\,\mb Q^2 - \frac{ \mc B}{3}\tr\,\mb Q^3 \right.\right. \nonumber \\ & \left.\left.+ \frac{1}{4}(\tr\,\mb Q^2)^2  - W \tr(\mb Q\mb X)\right] \right\}dz.
\end{align}

For spatially inhomogeneous systems, neither $\mb Q$ nor $\mb X$ can be simplified as in~\eqref{Q} and~\eqref{X}.  Following~\cite{sonnet1995alignment}, we represent an arbitrary $\mb Q$ tensor by
\begin{equation}
    \mb Q =  \frac{1}{\sqrt 2}\begin{bmatrix}
        -q_1/\sqrt 3 + q_2 & q_3 & q_4 \\ q_3 & -q_1/\sqrt 3-q_2 & q_5 \\ q_4 & q_5 & 2q_1/\sqrt 3
    \end{bmatrix}
\end{equation}
involving five degrees of freedom, $q_i=q_i(z)$, $i=1,2,3,4,5$. Similarly, the prescribed $\mb X$ tensor can be expressed using five constant values $x_i$. In this representation, the free energy simplifies to
\begin{align}
    \mc F[q_i] = \int_0^1 &\left\{\frac{1}{4}\left(\frac{4q_1'^2}{3} + q_4'^2+q_5'^2\right) + \frac{\eta}{2}\left[\frac{q_1'^2}{3} + q_2'^2 + q_3'^2 + \frac{q_4'^2}{2} + \frac{q_5'^2}{2} + \sigma (2q_3q_2'   -2q_2q_3' \right.\right.\nonumber \\
    &\left.\left. + q_5 q_4'-q_4q_5')  + \sigma^2\tr\,\mb Q^2\right] + \lambda\left[\frac{\mc A+\Sigma_s W}{2} \tr\,\mb Q^2 - \frac{ \mc B}{3}\tr\,\mb Q^3 + \frac{1}{4}(\tr\,\mb Q^2)^2  - W \tr(\mb Q\mb X) \right]\right\}dz, \nonumber
\end{align}
where the prime denotes differentiation with respect to $z$, $\tr\,\mb Q^2 = q_i q_i$, $\tr\,(\mb Q\mb X)= q_i x_i$ and $\tr\,\mb Q^3  = \tfrac{1}{2\sqrt 6}[2q_1^3-3q_1(2q_2^2+2q_3^2-q_4^2-q_5^2) + 3\sqrt 3 (q_2q_4^2-q_2q_5^2 +2 q_3 q_4 q_5)]$. The critical points and the minimizers of $\mc F[q_i]$ are classical solutions of the associated Euler-Lagrange (EL) equations:
\begin{align}
   (2+\eta)\frac{ q_1''}{3}   - \eta \sigma^2q_1  &= \lambda\left[ (\mc A +\Sigma_s W + \tr\,\mb Q^2)q_1 -\mc B\alpha_1 -  W x_1\right], \label{q1}\\
    \eta q_2''+2\eta\sigma q_3' - \eta\sigma^2 q_2 & = \lambda \left[(\mc A+\Sigma_s W + \tr\,\mb Q^2)q_2 -\mc B\alpha_2 -  W x_2\right], \label{q2}\\
  \eta q_3'' -2\eta\sigma q_2' - \eta\sigma^2 q_3 & = \lambda \left[(\mc A +\Sigma_s W+ \tr\,\mb Q^2)q_3 -\mc B\alpha_3 - W x_3\right] ,  \label{q3}\\
   (1+\eta)\frac{q_4''}{2} + \eta\sigma q_5' - \eta\sigma^2 q_4 &= \lambda \left[(\mc A +\Sigma_s W + \tr\,\mb Q^2)q_4 -\mc B\alpha_4 - W x_4\right] ,  \\
    (1+\eta)\frac{q_5''}{2} - \eta\sigma q_4' - \eta\sigma^2 q_5 &= \lambda \left[(\mc A +\Sigma_s W + \tr\,\mb Q^2)q_5 -\mc B\alpha_5 - W x_5\right] ,\label{q5}
\end{align}
where $3\alpha_i=\pfi{\tr\, \mb Q^3}{q_i}$, i.e., $2\sqrt 6\alpha_1 = 2(q_1^2-q_2^2-q_3^2)+q_4^2+q_5^2$, $2\sqrt 6\alpha_2=-4q_1q_2 +\sqrt 3(q_4^2-q_5^2)$, $2\sqrt 6\alpha_3=-4q_1q_3 + 2\sqrt 3 q_4 q_5$, $2\sqrt 6\alpha_4=2q_1q_4 + 2\sqrt 3(q_2q_4+q_3q_5)$ and $2\sqrt 6\alpha_5=2q_1q_5 + 2\sqrt 3(q_3q_4-q_2q_5)$. These five equations are solved subject to the Dirichlet boundary conditions which translate to the following conditions for $q_1 \ldots q_5$:
\begin{equation}
    q_1 = -\frac{s_+}{\sqrt 6}, \quad q_2 = \frac{s_+}{\sqrt 2}, \quad q_3=q_4=q_5=0 \quad \text{at}\quad z=0,1.  \label{dirichlet}
\end{equation}

If $\mb e_z$ is an eigenvector of $\mb X$ so that $x_4=x_5=0$, then there are solution branches with $q_4=q_5=0$~\cite{dalby2023order}. The stable solutions  typically correspond to these branches, whose eigenvalues $(\lambda_i)$ and eigenvectors $(\mb v_i$) can be written as
\begin{equation}
    \lambda_{1,2}= -\frac{q_1}{\sqrt 6} \pm \frac{\sqrt{q_2^2+q_3^2}}{\sqrt 2},\quad \lambda_3 = \frac{2q_1}{\sqrt 6}, \quad \mb v_{1,2}=\frac{1}{\mc N}\begin{bmatrix}
        q_2 \pm \sqrt{q_2^2+q_3^2} \\ q_3 \\ 0
    \end{bmatrix},\quad \mb v_3 =\mb e_z. \label{leadingeig}
\end{equation}
where $\mc N = \sqrt{2(q_2^2+q_3^2) \pm 2q_2\sqrt{q_2^2+q_3^2}}$. Thus, the planar eigenvectors depend only on $q_2$ and $q_3$ for solution branches with $\mb e_z$ as a fixed eigenvector. We focus on such solution branches in the remainder of this section.

\subsection{Thin-film limit asymptotics}
\label{sec:asymptotic}
Consider the thin-film limit, $\lambda \to 0$, while simultaneously maintaining a small pitch such that $\sigma\sim 1$ and a non-negligible influence from the NPs, i.e., $\lambda W\sim 1$. This limit explores the role of suspended NPs on purely elastic systems, complementing the analysis in Section~\ref{sec:homogeneous}, which focuses on homogeneous systems. The Euler--Lagrange equations~\eqref{q1}-\eqref{q5} simplify to 
\begin{align}
    \frac{2+\eta}{3\eta}q_1''   - ( \sigma^2+\gamma^2)q_1  &=-\varpi x_1, \label{firstq} \\ q_2''+ 2 \sigma q_3' - ( \sigma^2+\gamma^2) q_2 &=-\varpi x_2, \quad q_3''- 2 \sigma q_2' - ( \sigma^2+\gamma^2) q_3 =-\varpi x_3, \label{secondq} \\ \frac{1+\eta}{2\eta} q_4'' + \sigma q_5'- ( \sigma^2+\gamma^2) q_4 &=-\varpi x_4, \quad \frac{1+\eta}{2\eta} q_5'' - \sigma q_4'- ( \sigma^2+\gamma^2) q_5 =-\varpi x_5 \label{thirdq}
\end{align}
where $\varpi=\lambda W/\eta$ and $\gamma=\sqrt{\lambda W\Sigma_s/\eta}$. 

The first equation~\eqref{firstq} is decoupled from the others and can be readily solved to give
\begin{equation}
    q_1 = \frac{\varpi x_1}{\sigma^2 + \gamma^2} - \left(\frac{\varpi x_1}{\sigma^2 + \gamma^2}+\frac{s_+}{\sqrt 6}\right) \frac{\cosh[(1-2z)\gamma_1/2]}{\cosh\gamma_1/2}, \quad \gamma_1 = \sqrt{\frac{3\eta (\sigma^2+\gamma^2)}{2+\eta}}. \label{q1a}
\end{equation}
The equations~\eqref{secondq} for $q_2$ and $q_3$ are coupled to each other as are the equations~\eqref{thirdq} for $q_4$ and $q_5$. Each pair of coupled equations can be combined into a single equation for the complex variables, $u = q_2 + i q_3$ and $v = q_4 + i q_5$, i.e.,
\begin{align}
    u'' - 2 i \sigma u' - (\sigma^2 + \gamma^2) u = -\varpi (x_2+ix_3), &\quad u (0)=u (1) = \frac{s_+}{\sqrt 2},\\
    \frac{1+\eta}{2\eta}v'' -  i \sigma v' - (\sigma^2 + \gamma^2) v = -\varpi (x_4+ix_5), &\quad v (0)=v (1) =0.
\end{align}
The solutions for $u$ and $v$ are given by
\begin{align}
    &u = \frac{\varpi (x_2+ix_3)}{\sigma^2+\gamma^2} - \left(\frac{\varpi (x_2+ix_3)}{\sigma^2 + \gamma^2}-\frac{s_+}{\sqrt 2}\right)\frac{\sinh[\gamma(1-z)] e^{i\sigma z} + \sinh(\gamma z)\, e^{-i\sigma(1-z)}}{\sinh\gamma}, \label{q2q3a}\\
    &v =  \frac{\varpi (x_4+ix_5)}{\sigma^2+\gamma^2}\left\{1 -   \frac{\sinh[\gamma_4(1-z)] e^{i\eta\sigma z/(1+\eta)} + \sinh(\gamma_4 z)\, e^{-i\eta\sigma(1-z)/(1+\eta)}}{\sinh\gamma_4}\right\}, \label{q4q5a}
\end{align}
where $\gamma_4 = \tfrac{\eta}{1+\eta}\sqrt{\sigma^2+2\gamma^2+2(\sigma^2+\gamma^2)/\eta}$. 

For a nematic system, we set $\sigma=0$ in~\eqref{q1a},\eqref{q2q3a} and~\eqref{q4q5a} and the solution simplifies to
\begin{equation}
    \mb Q = \frac{\mb X}{\Sigma_s} + \left(-\frac{\mb X}{\Sigma_s} + \mb Q_0\right) \frac{\cosh[(1-2z)\gamma/2]}{\cosh\gamma/2}
\end{equation} in the one-constant approximation with $\eta=1$.

\section{Numerical results}
\label{sec:numerical}

 In this section, we present some numerical examples of structural transitions in dilute suspensions of colloidal NPs in cholesteric-filled channel geometries as described above. In particular, we show that the NPs can drive unwinding or untwisting transitions for cholesterics, driven by the imposed symmetry of $\mb X$.
 We work with a fixed prescribed $\mb X$ given by
\begin{equation} \label{fixedX}
    \mb X =\left( \mb e_x\otimes\mb e_x -\tfrac{1}{3}\mb I\right).
\end{equation}
Furthermore, we fix the parameters to be $\eta=4.5$, $\sigma=4\pi$ (as used in~\cite{han2022uniaxial}) and $\mc B=1.5$, $W=0.5$ and $\Sigma_s=4\pi$ (as used in Section~\ref{sec:homogeneous}). Two representative values for $\lambda$ are used for the illustrative examples: $\lambda=1000$ (corresponding to a relatively large domain, $H=0.25\mu m$ for $L_2=15pN$ and $C=2.4\times 10^5N/m^2$) and $\lambda=10$ (corresponding to a relatively small domain, $H=25nm$). The temperature parameter $\mc A$ is varied continuously. 

When $\mb X = \mb 0$ (or $W=0$) and in the $\lambda\to\infty$ limit, the minimizers of \eqref{cholesteric_energy} are well-approximated by uniaxial, helical $\mb Q$-fields of the form 
\begin{equation}
    \mb Q_\omega = s_+\left(\mb n_\omega\otimes\mb n_\omega -\tfrac{1}{3}\mb I\right), \quad \mb n_\omega = (\cos(\omega\pi z),\sin(\omega\pi z),0), \quad \omega \in \mathbb Z, \label{helical}
\end{equation}
which are splay/bend free ($\nabla\cdot\mb Q_\omega=0$, i.e., $q_1'=q_4'=q_5'=0$)~\cite{han2022uniaxial,dalby2023order}. The integer $\omega$ measures the number of $\pi$-rotations of the director field $\mb n_\omega$ across the domain. Accordingly, we refer to $\mb Q_1$ as a $\pi$-twisting configuration and $\mb Q_2$ as a $2\pi$-twisting configuration, etc. Among these, the $\mb Q_2$ configuration is typically the energy minimizer~\cite{han2022uniaxial}. 

Next, we present some heuristics for the case $\mb X \neq 0$. In the $\lambda \to \infty$ limit, the minimizers of~\eqref{cholesteric_energy} converge to the minimizer of the modified bulk potential $f_m$ in~\eqref{modified}, almost everywhere~\cite{majumdar2010landau}. The minimizers of $f_m$ converge to the minimizers of $f_B$ as $\mc A \to -\infty$, for fixed $W$, i.e., uniaxial $\mb Q$ of the form \eqref{eq:fbmin} with arbitrary $\mb n$. In this case, informally speaking, we minimize the cholesteric free energy in the restricted class of $\mb Q$-tensors in \eqref{eq:fbmin} and recover the helical $\mb Q$-fields in \eqref{helical}. However, for modestly large values of $\mc A$ and fixed $W$, the minimizers of $f_m$ (for the particular choice of $\mb X$ in \eqref{fixedX}; also see Figure~\ref{fig:prolateX})  instead converge to the uniaxial state described by
\[
\mb Q^* = S_+^x (\mb e_x \otimes \mb e_x - \tfrac{1}{3} \mb I)
\]
for sufficiently large values of $\lambda$. This motivates the untwisting transitions for large $\lambda$, as the temperature parameter $\mc A$ is increased, and we expect to see sharp transitions around $\mc A = -5$ since there is only one stable critical point, $(S_+^x,0)$, of $f_m$ for $\mc A >-5$, as plotted in Figure~\ref{fig:prolateX}.

\begin{figure}[h!]
\hspace{-0.8cm}
\includegraphics[width=0.38\textwidth]{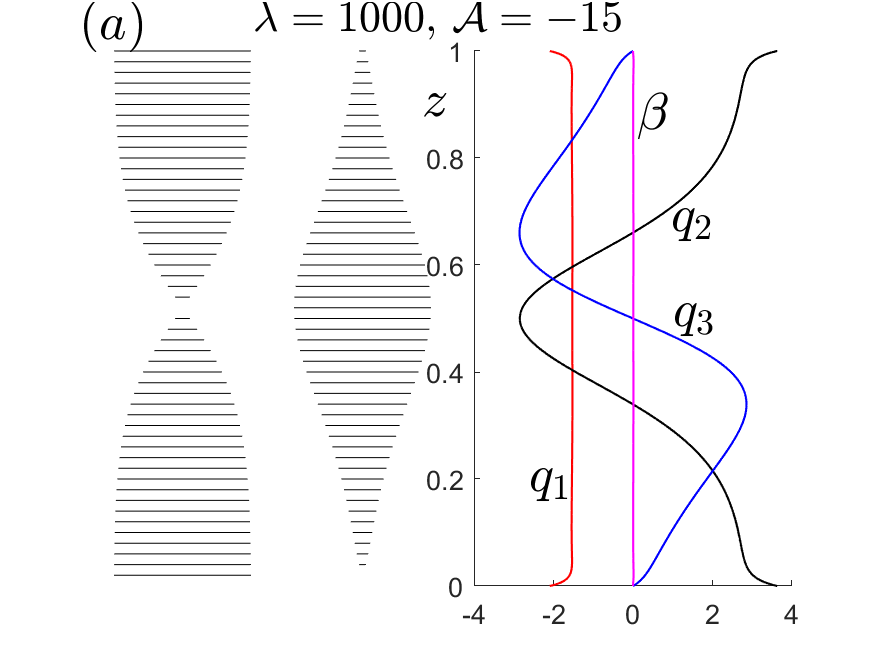}
\hspace{-0.8cm}
\includegraphics[width=0.38\textwidth]{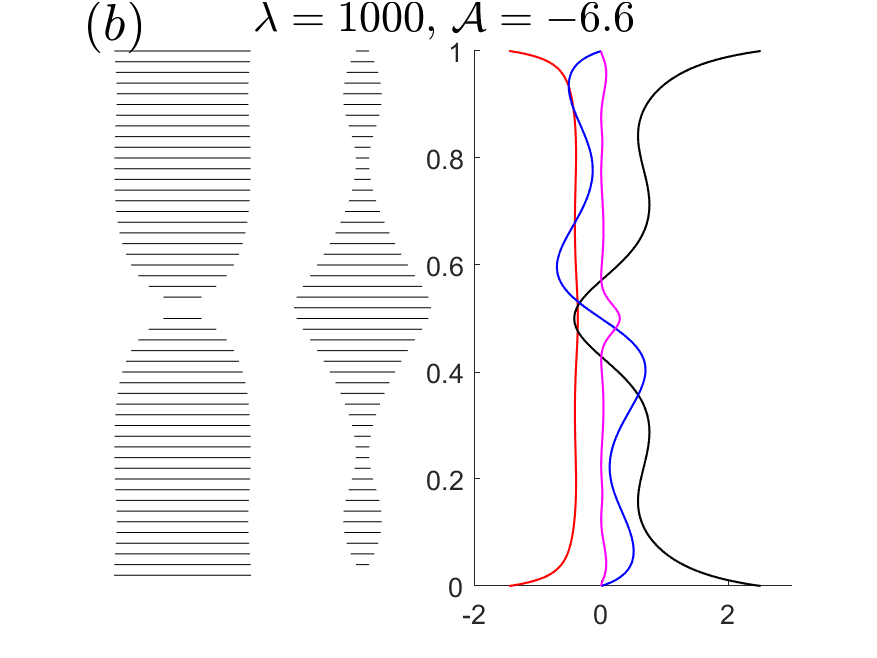}
\hspace{-0.8cm}
\includegraphics[width=0.38\textwidth]{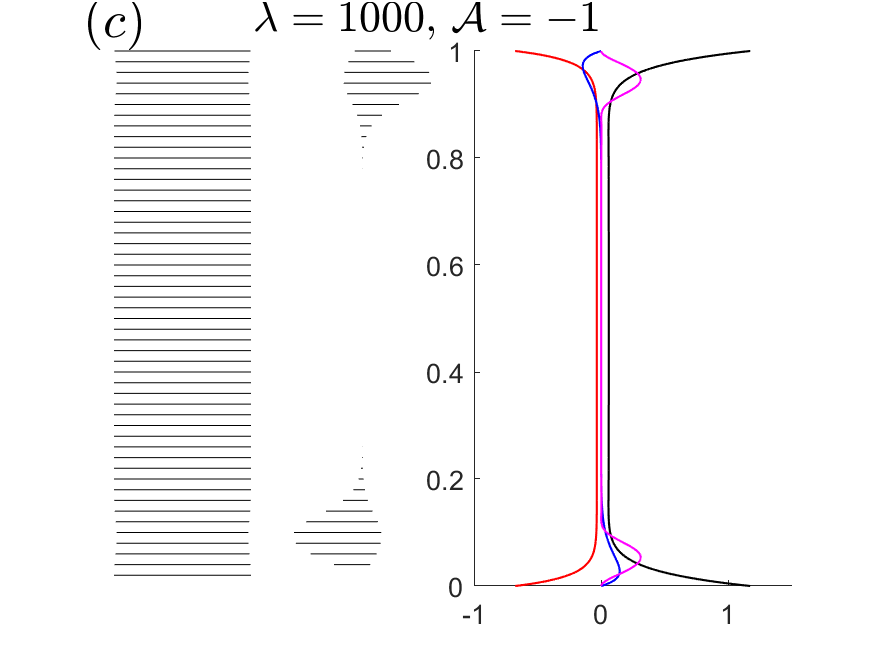}
\medskip
\hspace{-0.8cm}
\includegraphics[width=0.38\textwidth]{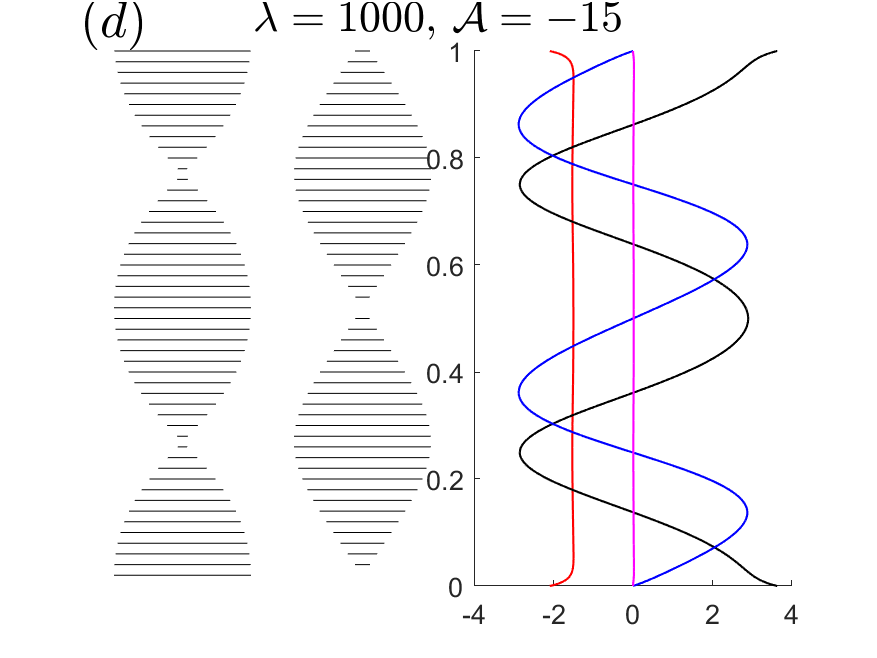}
\hspace{-0.8cm}
\includegraphics[width=0.38\textwidth]{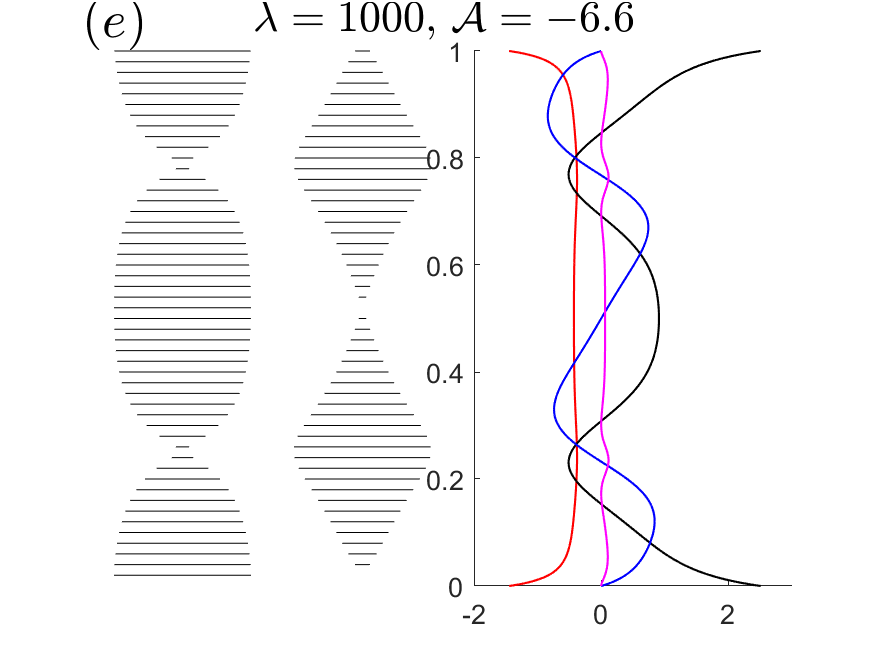}
\hspace{-0.8cm}
\includegraphics[width=0.38\textwidth]{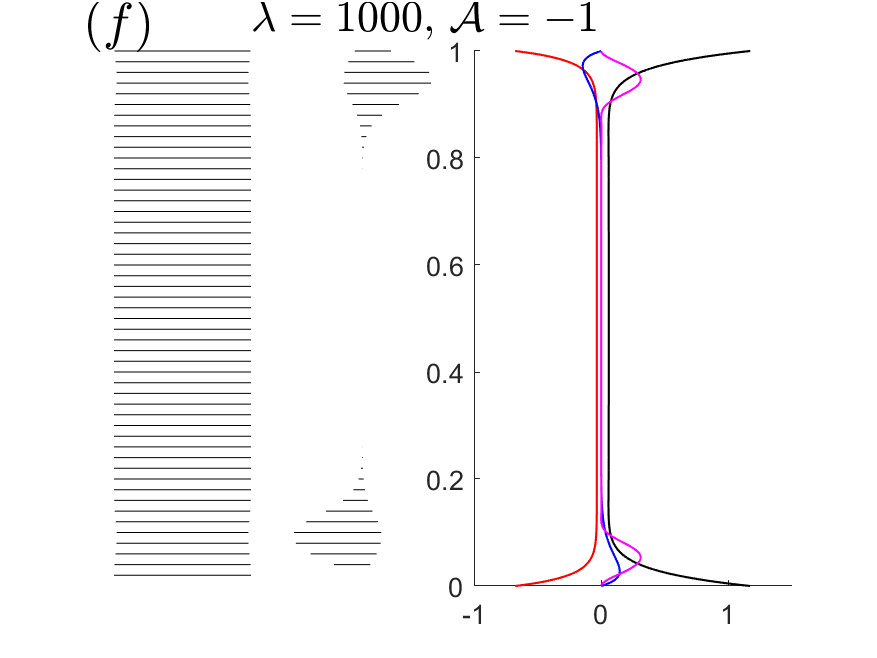}
\caption{Numerical solutions of \eqref{q1}-\eqref{q5} subject to the Dirichlet conditions in \eqref{dirichlet} that demonstrate the transition from the uniaxial helical $\mb Q$-field to an uniaxial (with thin biaxial boundary layers at $z=0,1$) untwisted $\mb Q$-field, as $\mc A$ is increased. The parameter values are $\lambda=1000$ with $W=0.5$, $\mc B=1.5$, $\eta=4.5$, $\sigma=4\pi$, $\Sigma_s=4\pi$ (same parametric values are used in subsequent figures) and $\mb X = \left(\mb e_x\otimes\mb e_x -\frac{1}{3}\mb I\right)$. The two snippets on the left in each subfigure shows the dominant director field  of $\mb Q$, projected onto $xz$- and $yz$-planes. The top row pertain to $\pi$-twisting helices and the bottom row to $2\pi$-twisting helices. In both cases, the helical $\mb Q$-fields disappear roughly around $\mc A=-5$ and this transition occurs at lower temperatures for larger $\lambda$.} 
\label{fig:largedomain}
\end{figure}

We numerically compute solutions of \eqref{q1}-\eqref{q5}, subject to the Dirichlet boundary conditions \eqref{dirichlet}, with $\mb X$ as in \eqref{fixedX}, and parametrically change $\mc A$ from $\mc A=-20$ to $\mc A=0$. \red{The computations are carried out using COMSOL Multiphysics, which employs the finite element method for spatial discretization and the MUMPS (Multifrontal Massively Parallel Sparse) direct solver. A mesh size of $\Delta z=10^{-3}$ is used after verifying mesh independence.} The results for $\lambda=1000$ are plotted in Figure~\ref{fig:largedomain} for the $\pi$-twisting and $2\pi$-twisting helices as initial guesses for the numerical solver. First of all, at a finite temperature (or at a finite $W$), the formula given in~\eqref{helical} is not strictly valid since a simple harmonic function is insufficient to characterise the solution, as can be inferred from the figure. However, the solutions are still found to be splay/bend free, except in small neighbourhoods of $z=0,1$. The plots for $\mc A=-6.6$ show that there is significant local biaxiality near the points $z_n=n/2\omega$, $n=1,3,5,\dots2\omega-1$. Both the $\pi$-twisting and $2\pi$-twisting solutions then disappear roughly at  $\mc A = -5$ and the solution transitions to the untwisted solution, dictated by the minimizer of $f_m$ for $\mc A = -5$ in Figure~\ref{fig:prolateX}. Notably, this untwisting transition becomes more abrupt and occurs at lower temperatures for larger values of $\lambda$, where the bulk effects (governed by $f_m$) dominate the energy landscape. For instance, when $\lambda=7000$, the untwisting occurs near $\mc A= -8$. We also perform computations by decreasing $\mc A$ parametrically, using  $\mb Q = \mb X/\Sigma_s$ as an initial guess. These suggest that the untwisted solution can persist to large negative values of $\mc A$, although we expect such solutions to be unstable at sufficiently low temperatures, due to the preference for twisted minimizers.

\begin{figure}[h!]
\hspace{-0.8cm}
\includegraphics[width=0.38\textwidth]{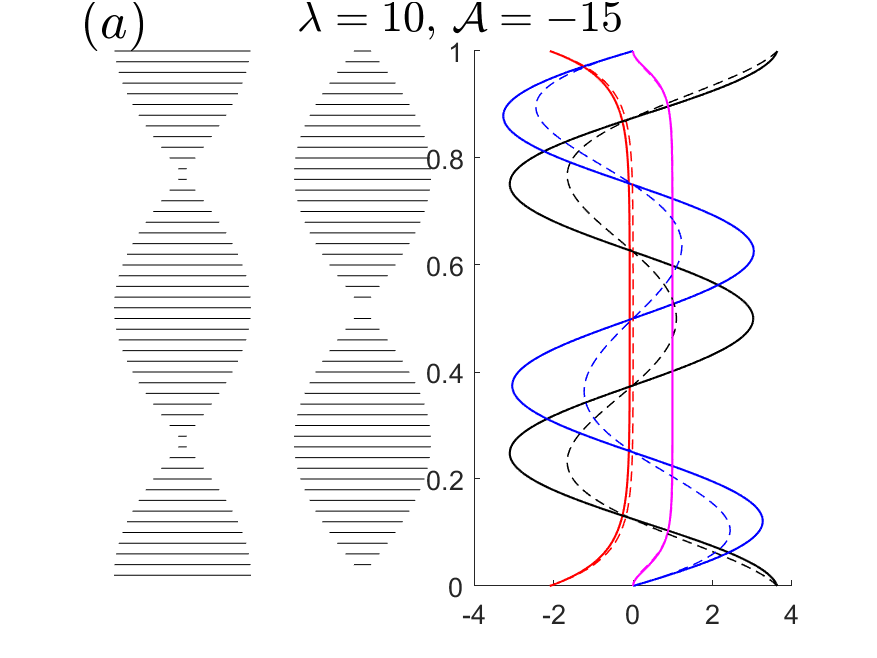}
\hspace{-0.8cm}
\includegraphics[width=0.38\textwidth]{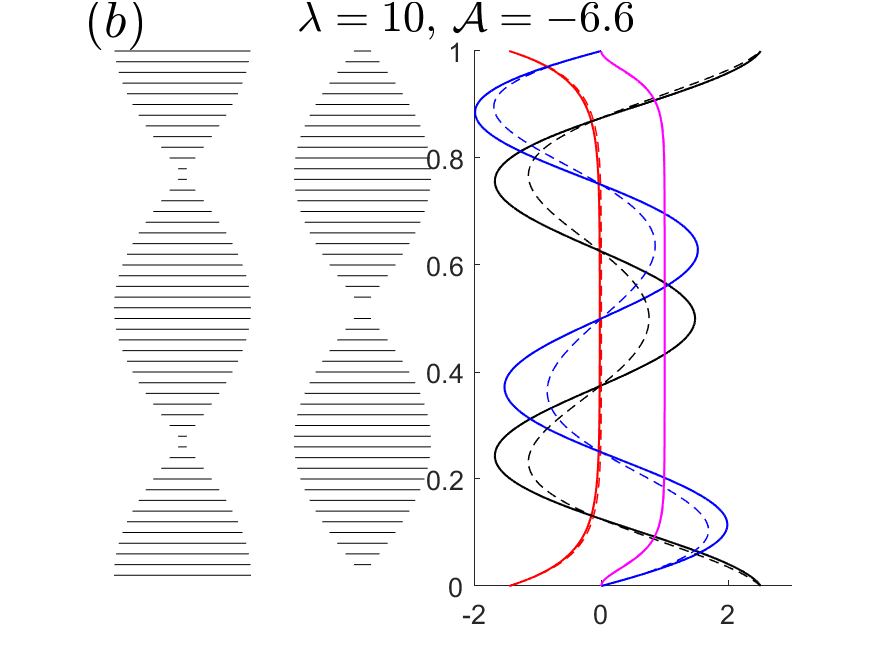}
\hspace{-0.8cm}
\includegraphics[width=0.38\textwidth]{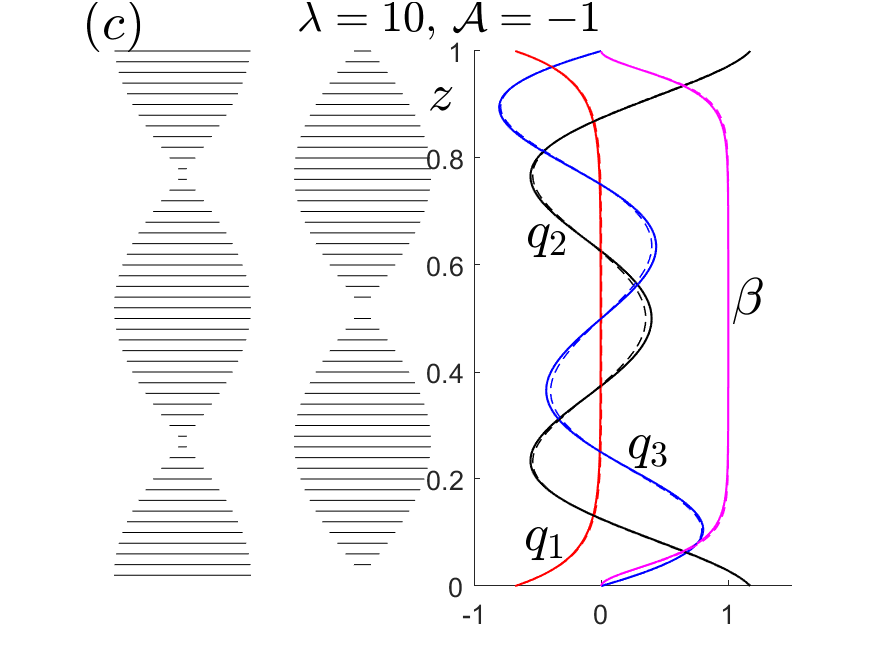}
\caption{Numerical solutions of \eqref{q1}-\eqref{q5} subject to the Dirichlet conditions in \eqref{dirichlet} with $\mb X = \left(\mb e_x\otimes\mb e_x -\frac{1}{3}\mb I\right)$ and $\lambda=10$. These solutions demonstrate a mild weakening of biaxial, helical configurations as $\mc A$ increases. For this value of $\lambda=10$, untwisting is expected to occur only for large positive values of $\mc A$. The dashed lines pertain to the asymptotic solution~\eqref{q1a}-\eqref{q2q3a}, which are applicable in the $\lambda \to 0$ limit with $|\mc A|\sim 1$. } 
\label{fig:smalldomain}
\end{figure}

Next, consider a much smaller domain corresponding to $\lambda=10$. We note that $\omega\pi$-twisting solutions with $\omega \neq 2$, are generally not energetically favourable for such small values of $\lambda$. In fact, the minimizer of the cholesteric free energy \eqref{cholesteric_energy} converges to the biaxial, $2\pi$-twisting solution in ~\eqref{q1a}-\eqref{q4q5a} in the $\lambda \to 0$ limit. The effect of changing $\mc A$ on these biaxial, $2\pi$-twisting solutions is illustrated in Figure~\ref{fig:smalldomain}. As $\mc A$ increases, the helical structure is gradually suppressed, although not completely eliminated in this regime. This indicates that the untwisting transition occurs for large positive values of $\mc A$, when $\lambda$ is small. In Figure~\ref{fig:smalldomain}, the asymptotic solutions for the limits $\lambda\to 0$ with $\mc A\sim 1$ in Section~\ref{sec:asymptotic}, are plotted as dashed lines for the purpose of comparison.  Although $\lambda =10$ is not a particularly small number, the agreement between the numerical solutions and the asymptotic solutions is satisfactory, especially for the value $\mc A=-1$.

\begin{figure}[h!]
\hspace{-0.8cm}
\includegraphics[width=0.38\textwidth]{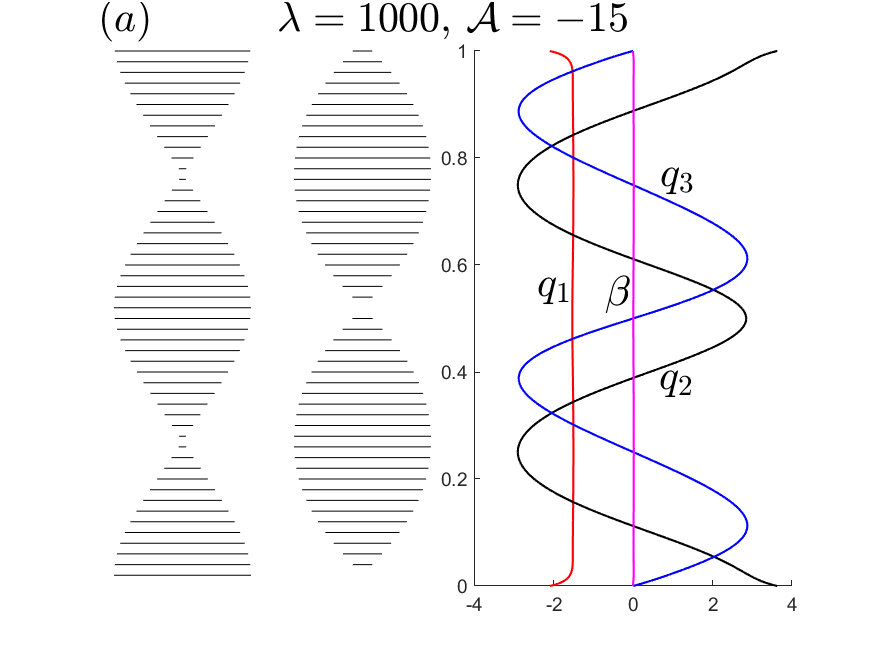}
\hspace{-0.8cm}
\includegraphics[width=0.38\textwidth]{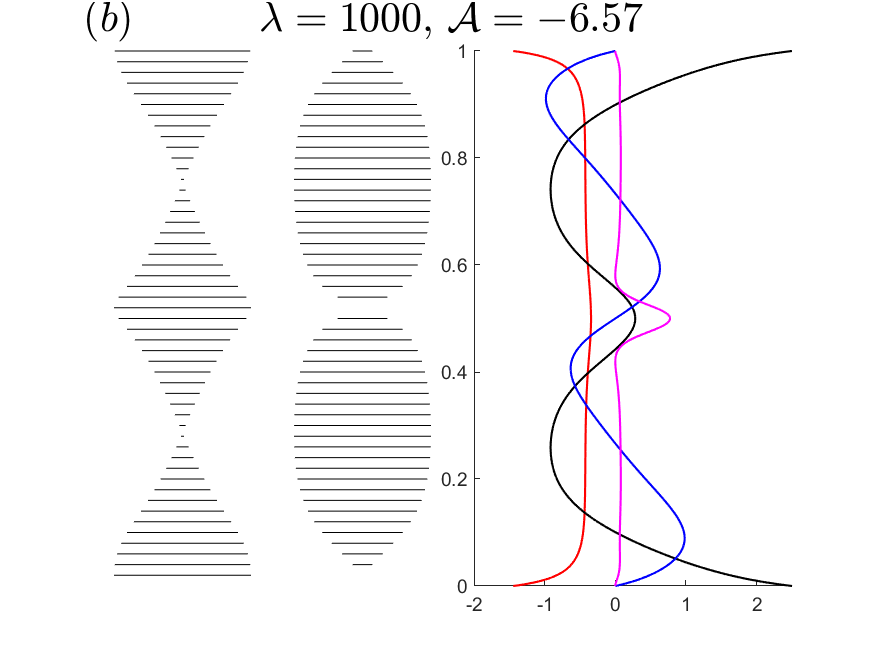}
\hspace{-0.8cm}
\includegraphics[width=0.38\textwidth]{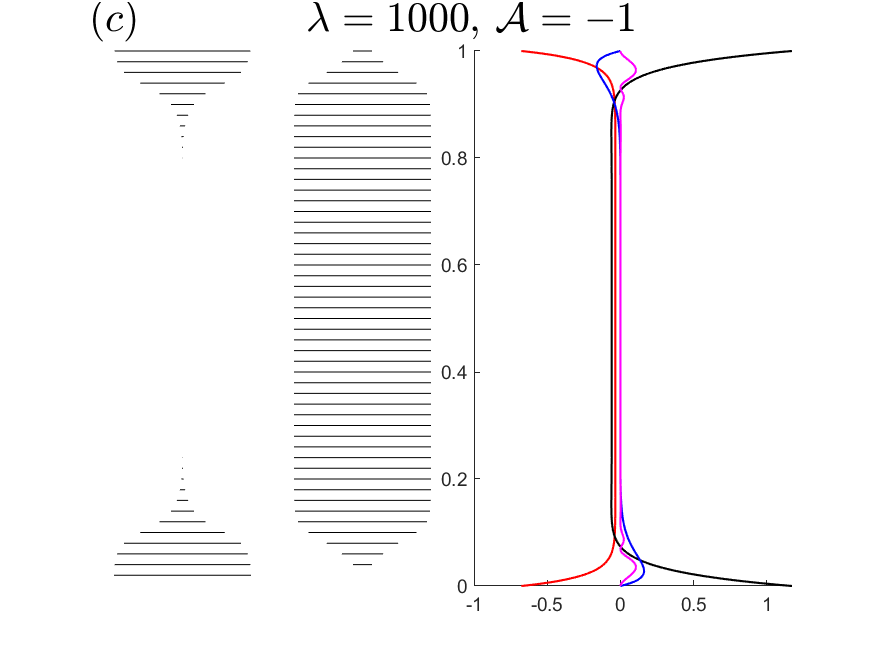}
\caption{Numerical solutions of \eqref{q1}-\eqref{q5}, subject to \eqref{dirichlet}, with $\mb X = \left(\mb e_y\otimes\mb e_y -\frac{1}{3}\mb I \right)$ and $\lambda=1000$. These solutions illustrate the transition from the uniaxial, $2\pi$-twisting helical $\mb Q$-field to  an uniaxial (with thin biaxial boundary layers at $z=0,1$), untwisted $\mb Q$-field, with increasing temperature.} 
\label{fig:helicaly}
\end{figure}

\begin{figure}[h!]
\hspace{-0.8cm}
\includegraphics[width=0.38\textwidth]{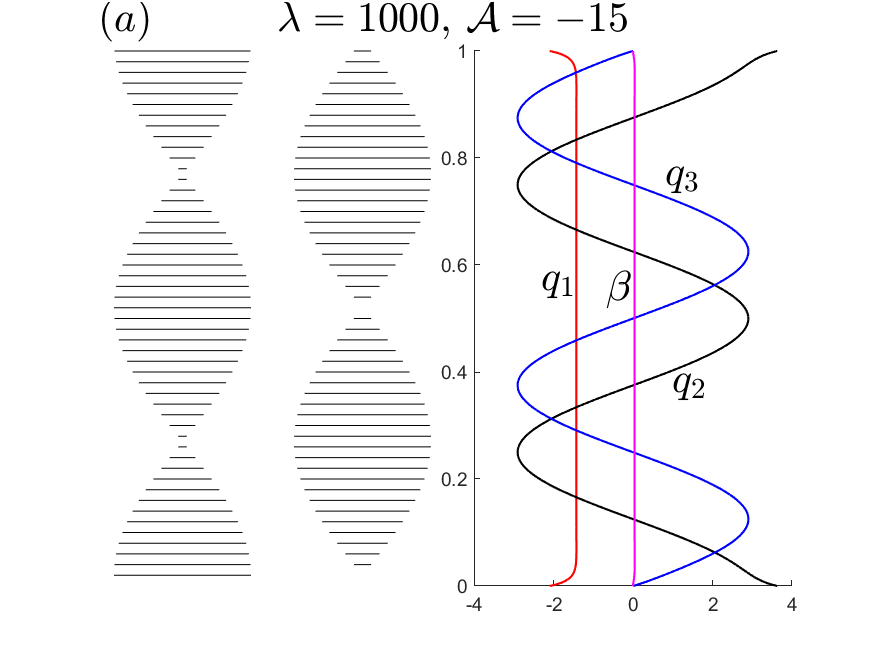}
\hspace{-0.8cm}
\includegraphics[width=0.38\textwidth]{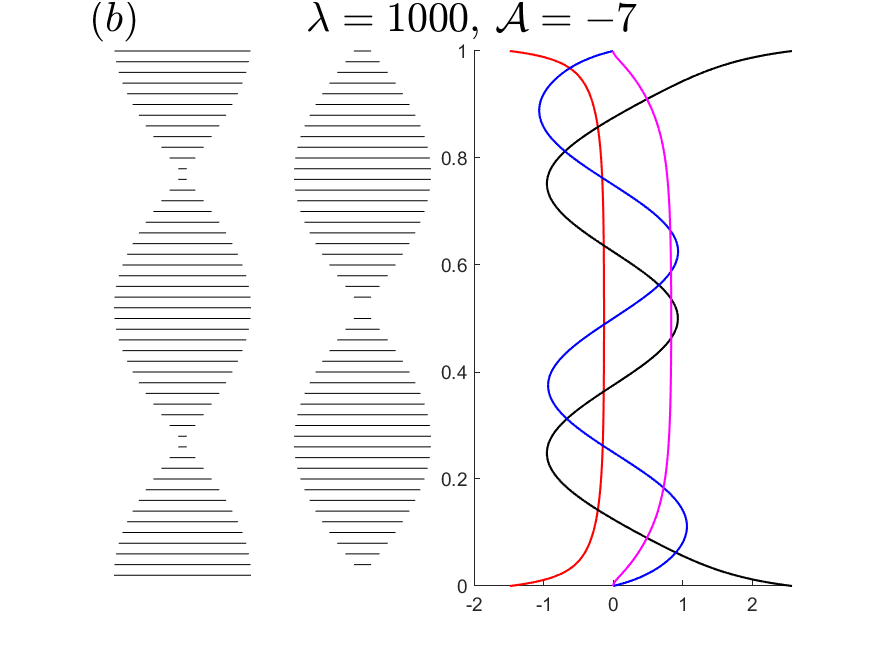}
\hspace{-0.8cm}
\includegraphics[width=0.38\textwidth]{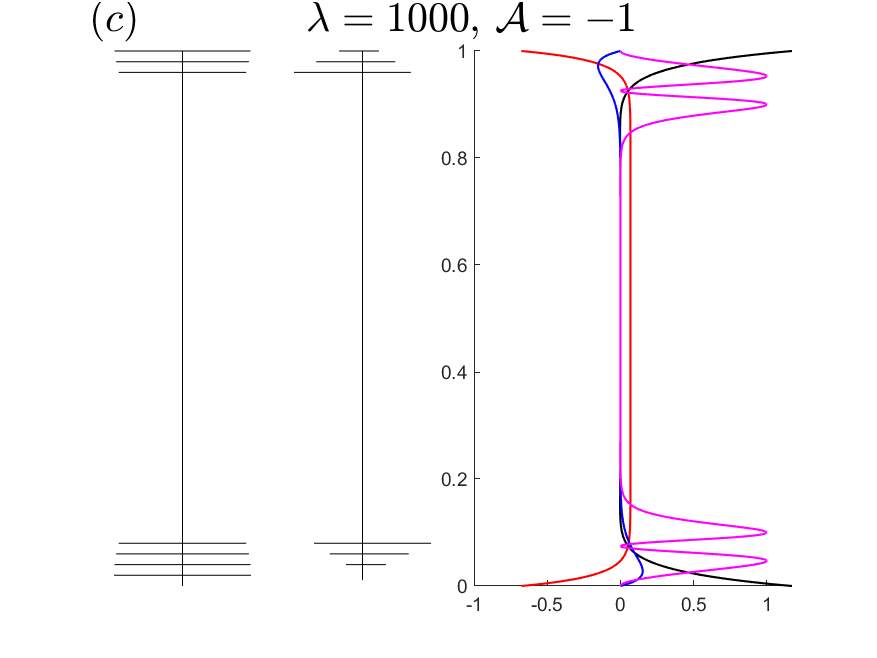}
\caption{Numerical solutions of \eqref{q1}-\eqref{q5}, subject to \eqref{dirichlet}, with $\mb X = \left(\mb e_z\otimes\mb e_z -\frac{1}{3}\mb I \right)$ and $\lambda=1000$. These solutions illustrate the transition from the uniaxial, $2\pi$-twisting helical $\mb Q$-field to an uniaxial (with thin biaxial boundary layers at $z=0,1$) untwisted $\mb Q$-field, with increasing temperature. \red{In the last case, the leading eigenvector is $\mb e_z$ almost everywhere and therefore its projections onto the $xz$- and $yz$-planes are null.} } 
\label{fig:helicalz}
\end{figure}

In Figure~\ref{fig:helicaly} and~\ref{fig:helicalz}, we study the untwisting transitions for the cases with
\begin{equation}
    \mb X = (\mb e_y \otimes \mb e_y - \tfrac{1}{3}\mb I) \quad \text{and} \quad  \mb X = (\mb e_z \otimes \mb e_z - \tfrac{1}{3}\mb I),
\end{equation}
respectively. The implications of these figures are qualitatively similar to that of the case with  $\mb X = (\mb e_x \otimes \mb e_x - \tfrac{1}{3}\mb I)$ in the sense that we get untwisting and lose the helical profiles as $\mc A$ increases, for $\lambda=1000$. There are negligible differences between Figure~\ref{fig:largedomain}(d) and Figure~\ref{fig:helicaly}(a), Figure~\ref{fig:helicalz}(a), since the stable solutions of \eqref{q1}-\eqref{q5} converge to uniaxial $\mb Q$-fields of the form \eqref{helical} for large and negative values of $\mc A$. As $\mc A$ increases, the stable solutions converge to \[
\mb Q = S_+^x (\mb e_y \otimes \mb e_y - \tfrac{1}{3}\mb I)
\]
for $\mb X = (\mb e_y \otimes \mb e_y - \tfrac{1}{3}\mb I)$ with $S_+^x$ as  plotted in Figure~\ref{fig:prolateX}, and similarly to
\[
\mb Q =  S_+^x(\mb e_z \otimes \mb e_z - \tfrac{1}{3}\mb I)
\] for $\mb X = (\mb e_z \otimes \mb e_z - \tfrac{1}{3}\mb I)$. In both cases, the helical symmetry is broken by the suspended NPs or by the leading director of $\mb X$. We get some peaks in the biaxiality near the edges, $z=0$ and $z=1$, induced by the boundary layers that arise from the conflicting boundary condition in \eqref{planar}.

To assess the stability of the stationary solutions discussed above, parallel time-dependent computations are performed using the gradient-flow equations
\begin{align}
    \pfr{q_i}{t} = -\frac{\delta \mc F}{\delta q_i},
\end{align}
where the right-hand side is effectively obtained from ~\eqref{q1}-\eqref{q5}. In the computations, the temperature parameter $\mc A$ is varied quasi-steadily while keeping all other parameters fixed. Specifically, we assume
\begin{equation}
    \mc A(t) = \mc A(0) + \Delta \mc A \left(\frac{1}{2} + \left\lfloor \frac{t}{\Delta t}\right\rfloor\right) + \frac{\tanh(\Delta t_w (t/\Delta t-\lfloor t/\Delta t\rfloor - 1/2))}{\tanh(\Delta t_w/2)} 
\end{equation}
with $\mc A(0)=-20$ (or $\mc A(0)=0$), $\Delta \mc A=+ 0.5$ ($\Delta \mc A = -0.5$), $\Delta t=100$ and $\Delta t_w=10$. \red{For time-dependent simulations, a relative tolerance of $10^{-3}$ and a fixed time step of $0.05$ are used in COMSOL Multiphysics.} This function slowly increases (or decreases) $\mc A$ in steps of $\Delta A$ across every $\Delta t$ time interval; the jump in $\mc A$ is smoothed out by the $\tanh$ functions with a transition width of $\Delta t_w$. The quasi-steady variation allows us to assess the stability characteristics of the solutions in Figures~\ref{fig:largedomain}-\ref{fig:helicalz}, obtained by means of stationary computations. We find that the computed stationary solutions are stable. For large negative values of $\mc A$, the helical solutions are stable whereas the untwisted solutions are unstable. Moreover, when $\mc A$ is gradually increased from its initial value $\mc A(0)=-20$, the helical solutions transition to the untwisted solutions at values of $\mc A$ close to those found in stationary computations. Similarly, when $\mc A$ is decreased from $\mc A(0)=0$, the reverse transition (from untwisted to helical) occurs at lower temperatures than the helical-untwisted transition, thereby indicating the presence of hysteresis. Such hysteresis behaviour in cholesteric LC samples, doped with quantum dots, has been previously observed in the experiments conducted in~\cite{rodarte2012spectral}. This also opens avenues for changing the optical properties of a doped cholesteric system by means of increasing or decreasing temperature as opposed to applying external fields. In particular, we break the director symmetry with the $\mb X$-field--generated by a dilute suspension of colloidal NPs--rather than by an external electric field. This will be investigated further in future work.

\section{Conclusions}
\label{sec:conclusions}

We have rigorously explored the influence of spherical colloidal NPs on  liquid crystals using the homogenized LdG framework as developed in \cite{canevari2020design}, in the dilute limit. For spatially homogeneous systems, the LdG bulk energy $f_B$ in \eqref{eq:2} is replaced by the modified bulk potential, $f_m$ in \eqref{modified}, to account for the collective effects of suspended spherical NPs with identical boundary treatments, in the dilute limit. The suspended NPs are fully characterised by $W$ (the  anchoring strength of the NP--LC interface), the NP geometry (spherical in this case), and the preferred boundary condition, $\mb Q_\nu$, on all NP-LC interfaces. The preferred boundary condition is imposed by means of a Rapini--Papoular surface energy \cite{canevari2020design}, although other choices are possible too. The modified bulk potential $f_m$ differs from $f_B$ through an effective temperature $\tilde{\mc A}$ and a linear energy contribution induced by a constant matrix $\mb X$, that originates from the homogenized contribution of the NP-LC surface energies summed over all NP-LC interfaces. The effective temperature increases for $W>0$, and the bulk symmetry is broken if $\mb X \neq 0$. Notably, we consider different possibilities for $\mb X$, corresponding to arbitrary $\mb Q_\nu$; in this respect, our study is perhaps the most general to date.  If $\mb X =0$, then the critical points of $f_m$ are either uniaxial or isotropic (as in the case of $f_B$) but the critical temperatures are shifted, i.e., we get a first-order isotropic-nematic phase transition at a lower temperature for $W>0$, as compared with the $W=0$ case. For $\mb X \neq 0$, we consider four representative cases: (i) uniaxial $\mb X$ with prolate or oblate symmetry and (ii) biaxial $\mb X$ with prolate or oblate symmetry. In all cases, we obtain seven critical points of $f_m$ which approach the critical points of $f_B$ for low temperatures (i.e., as $\mc A \to -\infty$). The global minimizer of $f_m$ always has prolate symmetry at low temperatures, as in the case of $f_B$. Furthermore, we always observe locally stable biaxial bulk phases induced by $\mb X$, for a certain range of temperatures, which are otherwise inaccessible  without the NPs. There is no first order isotropic-nematic phase transition when $\mb X \neq 0$ and we instead observe a gradual crossover between ordered and disordered states, with increasing temperature. At high temperatures, there is a unique minimizer of $f_m$ that inherits the prolate or oblate symmetry of $\mb X$, again a behaviour that is not possible without the NPs. Our work thus highlights the capacity of NPs to stabilize phases that are otherwise inaccessible in pure nematic or cholesteric systems.  

We complement our work on spatially homogeneous samples with a simple example of a dilute suspension of spherical NPs in a cholesteric-filled channel geometry, subject to planar Dirichlet boundary conditions on the the top and bottom surfaces. This is a well-studied example and people expect to find stable twisted cholesteric director profiles, when the channel is sufficiently large, i.e., when the confinement parameter $\lambda$ is sufficiently large. However, this symmetry is broken by the suspended NPs, with increasing temperature. We demonstrate structural transitions from stable cholesteric helices to almost uniform director profiles with increasing temperature, provided  $\lambda$ is  sufficiently large. The uniform director profiles inherit the symmetry of $\mb X$ when $\mc A$ and $\lambda$ are large enough. Equally importantly, this transition is reversible, i.e., we can numerically observe untwisted - twisted helical transitions induced by slowly decreasing the temperature, for sufficiently large $\lambda$. While this is a specific example, the qualitative conclusions will remain unchanged for generic confined cholesteric or nematic systems. Such systems tend to exhibit multistablility in large geometries (i.e., with large $\lambda$) under suitably prescribed boundary conditions. In other words, there are typically multiple LdG energy minimizers corresponding to multiple physically observable functional modes of the large LC system. This multistability will typically be lost when $\mb X \neq 0$, since the NPs impose a unique director orientation (specified by the symmetry of $\mb X$), for sufficiently large $\mc A$ and $\lambda$. An especially interesting example would be a cholesteric or nematic-filled three-dimensional spherical droplet with radial boundary conditions \cite{mclauchlan2024radial, sonnet1995alignment}, which naturally hosts interior defects, as a consequence of non-trivial topology. The radial symmetry can be broken by the suspended NPs in a spherical droplet, if $\mb X \neq 0$, and it would be interesting to investigate how $\mb X$ tailors interior defect structures in such confined systems in future work. 

Overall, our findings underscore the potential of NPs as a versatile tool to engineer novel phases and structural transitions in liquid crystals. We believe that our conclusions will remain valid for general NP shapes (other than spheres) and also for generic surface energies, within the dilute limit. For example, in \cite{canevari2020design, canevari2020polydispersity}, the authors consider a generic surface energy density on the NP-LC interface given by
\[
f_s(\mb Q, \nu) = f_s(\textrm{tr}\mb Q^2, \textrm{tr}\mb Q^3, \mb Q \gb \nu \cdot\gb \nu, \mb Q^2 \gb \nu \cdot \gb \nu)
\]
where $\gb \nu$ is the outward normal to the NP surface. One can directly show that post homogenization, such generic NP-LC surface energies only alter the coefficients of the original bulk energy, $f_B$ in \eqref{eq:2} and potentially add a linear contribution of the form, $W \textrm{tr}\left(\mb Q \mb X\right)$, where $\mb X$ encodes the geometric and anchoring characteristics of the NPs. Given that our analysis holds for arbitrary $\mb X$, our results are expected to largely carry over to more general scenarios. On these grounds, we speculate that our theoretical insights can be useful for future experiments on tunable, responsive materials with customized optical and mechanical properties.


\section*{Acknowledgments}
A.M. gratefully acknowledges support from the Leverhulme Research Project Grant RPG-2021-401 and an Isaac Newton Institute Network Grant, EPSRC Grant EP/R014604/1.
P.R. is supported by the Leverhulme Research Project Grant RPG-2021-401. This work was completed whilst A.M. was working at the University of Strathclyde.

\bibliographystyle{plain}
\bibliography{reference}

\begin{thebibliography}{10}

\bibitem{alexander2006stabilizing}
G.~P. Alexander and J.~M. Yeomans.
\newblock Stabilizing the blue phases.
\newblock {\em Phys. Rev. E}, 74(6):061706, 2006.

\bibitem{basu2009evidence}
R.~Basu and G.~S. Iannacchione.
\newblock Evidence for directed self-assembly of quantum dots in a nematic liquid crystal.
\newblock {\em Phys. Rev. E}, 80(1):010701, 2009.

\bibitem{batchelor1972sedimentation}
G.~K. Batchelor.
\newblock Sedimentation in a dilute dispersion of spheres.
\newblock {\em J. Fluid Mech.s}, 52(2):245--268, 1972.

\bibitem{bennett2017multiscale}
T.~Bennett.
\newblock {\em Multiscale modelling and experimental estimation of liquid crystals parameters}.
\newblock PhD thesis, University of Southampton, 2017.

\bibitem{bennett2018multiscale}
T.~P. Bennett, G.~D'Alessandro, and K.~R. Daly.
\newblock Multiscale models of metallic particles in nematic liquid crystals.
\newblock {\em SIAM J. Appl. Math.}, 78(2):1228--1255, 2018.

\bibitem{bisoyi}
H.~K. Bisoyi and Q.~Li.
\newblock Liquid crystals: Versatile self-organized smart soft materials.
\newblock {\em Chem. Rev.}, 122(5):4887--4926, 2022.

\bibitem{canevari2020design}
G.~Canevari and A.~Zarnescu.
\newblock Design of effective bulk potentials for nematic liquid crystals via colloidal homogenisation.
\newblock {\em Math. Models Methods Appl. Sci.}, 30(02):309--342, 2020.

\bibitem{canevari2020polydispersity}
G.~Canevari and A.~Zarnescu.
\newblock Polydispersity and surface energy strength in nematic colloids.
\newblock {\em Math. Eng.}, 2(2):290--312, 2020.

\bibitem{dalby2023order}
J.~Dalby.
\newblock Order reconstruction and solution landscapes for liquid crystalline systems.
\newblock 2023.

\bibitem{de1971short}
P.~G. de~Gennes.
\newblock Short range order effects in the isotropic phase of nematics and cholesterics.
\newblock {\em MCLC}, 12(3):193--214, 1971.

\bibitem{dg}
P.~G. de~Gennes and J.~Prost.
\newblock {\em {The Physics of Liquid Crystals}}, volume~83.
\newblock Oxford University Press, 1995.

\bibitem{fukuda2010cholesteric}
J.~Fukuda and S.~{\v{Z}}umer.
\newblock Cholesteric blue phases: effect of strong confinement.
\newblock {\em Liq. Cryst.}, 37(6-7):875--882, 2010.

\bibitem{han2022uniaxial}
Y.~Han, J.~Dalby, A.~Majumdar, B.~M. G.~D. Carter, and T.~Machon.
\newblock Uniaxial versus biaxial pathways in one-dimensional cholesteric liquid crystals.
\newblock {\em Phys. Rev. Res.}, 4(3):L032018, 2022.

\bibitem{kinkead2010effects}
B.~Kinkead and T.~Hegmann.
\newblock Effects of size, capping agent, and concentration of cdse and cdte quantum dots doped into a nematic liquid crystal on the optical and electro-optic properties of the final colloidal liquid crystal mixture.
\newblock {\em Journal of Materials Chemistry}, 20(3):448--458, 2010.

\bibitem{kumar2018quantum}
J.~Kumar, V.~Prasad, and M.~Manjunath.
\newblock Quantum dots dispersed hockey stick nematic liquid crystal: studies on dielectric permittivity, elastic constants and electrical conductivity.
\newblock {\em Journal of Molecular Liquids}, 266:10--18, 2018.

\bibitem{lagerwallnature}
X.~Ma, Y.~Han, Y.-S. Zhang, Y.~Geng, A.~Majumdar, and J.~P.~F. Lagerwall.
\newblock Tunable templating of photonic microparticles via liquid crystal order-guided adsorption of amphiphilic polymers in emulsions.
\newblock {\em Nat. Commun.}, 15(1):1404, 2024.

\bibitem{majumdar2010equilibrium}
A.~Majumdar.
\newblock Equilibrium order parameters of nematic liquid crystals in the {L}andau-de {G}ennes theory.
\newblock {\em Eur. J. Appl. Math.}, 21(2):181--203, 2010.

\bibitem{majumdar2010landau}
A.~Majumdar and A.~Zarnescu.
\newblock Landau--de gennes theory of nematic liquid crystals: the oseen--frank limit and beyond.
\newblock {\em Archive for rational mechanics and analysis}, 196:227--280, 2010.

\bibitem{mclauchlan2024radial}
S.~McLauchlan, Y.~Han, M.~Langer, and A.~Majumdar.
\newblock The radial hedgehog solution in the landau--de gennes theory: Effects of the bulk potentials.
\newblock {\em Physica D: Nonlinear Phenomena}, 459:134019, 2024.

\bibitem{mirzaei2012quantum}
J.~Mirzaei, M.~Reznikov, and T.~Hegmann.
\newblock Quantum dots as liquid crystal dopants.
\newblock {\em Journal of Materials Chemistry}, 22(42):22350--22365, 2012.

\bibitem{mirzaei2011nanocomposites}
J.~Mirzaei, M.~Urbanski, K.~Yu, H.~Kitzerow, and T.~Hegmann.
\newblock Nanocomposites of a nematic liquid crystal doped with magic-sized cdse quantum dots.
\newblock {\em Journal of Materials Chemistry}, 21(34):12710--12716, 2011.

\bibitem{mitov2017cholesteric}
M.~Mitov.
\newblock Cholesteric liquid crystals in living matter.
\newblock {\em Soft matter}, 13(23):4176--4209, 2017.

\bibitem{mottram2014introduction}
N.~J. Mottram and C.~J.~P. Newton.
\newblock Introduction to q-tensor theory.
\newblock {\em arXiv preprint arXiv:1409.3542}, 2014.

\bibitem{patranabish2021quantum}
S.~Patranabish, Y.~Wang, A.~Sinha, and A.~Majumdar.
\newblock Quantum-dots-dispersed bent-core nematic liquid crystal and cybotactic clusters: experimental and theoretical insights.
\newblock {\em Phys. Rev. E}, 103(5):052703, 2021.

\bibitem{rodarte2012spectral}
A.L. Rodarte, C.~Gray, L.S. Hirst, and S~Ghosh.
\newblock Spectral and polarization modulation of quantum dot emission in a one-dimensional liquid crystal photonic cavity.
\newblock {\em Phys. Rev. B Condens. Matter}, 85(3):035430, 2012.

\bibitem{singh2016emissivity}
G.~Singh, M.~Fisch, and S.~Kumar.
\newblock Emissivity and electrooptical properties of semiconducting quantum dots/rods and liquid crystal composites: a review.
\newblock {\em Reports on Progress in Physics}, 79(5):056502, 2016.

\bibitem{sonnet1995alignment}
A.~Sonnet, A.~Kilian, and S.~Hess.
\newblock Alignment tensor versus director: Description of defects in nematic liquid crystals.
\newblock {\em Phys. Rev. E}, 52(1):718, 1995.

\bibitem{stephen1974physics}
M.~J. Stephen and J.~P. Straley.
\newblock Physics of liquid crystals.
\newblock {\em Rev. Mod. Phys.}, 46(4):617, 1974.

\bibitem{lagerwallnew}
M.~Urbanski, C.~G. Reyes, J.~Noh, A.~Sharma, Y.~Geng, V.~S.~R. Jampani, and J.~P. Lagerwall.
\newblock Liquid crystals in micron-scale droplets, shells and fibers.
\newblock {\em J. Phys. Condens. Matter}, 29(13):133003, 2017.

\bibitem{wright1989crystalline}
D.~C. Wright and N.~D. Mermin.
\newblock Crystalline liquids: the blue phases.
\newblock {\em Rev. Mod. Phys.}, 61(2):385, 1989.

\bibitem{zhang2009cds}
T.~Zhang, C.~Zhong, and J.~Xu.
\newblock Cds-nanoparticle-doped liquid crystal displays showing low threshold voltage.
\newblock {\em Japanese Journal of Applied Physics}, 48(5R):055002, 2009.

\end{thebibliography}


\end{document}